\documentclass[12pt]{iopart}

\pdfoutput=1

\usepackage[english]{babel}
\usepackage{
amssymb,amsfonts,latexsym}
\usepackage{graphicx}
\usepackage{color}
\usepackage{iopams}
\usepackage{subfigure}
\usepackage{afterpage}
\usepackage{cite}
\usepackage{tikz}
\usetikzlibrary{snakes}
\usetikzlibrary{calc}
\usepackage{bm}

\definecolor{blue(pigment)}{rgb}{0.2, 0.2, 0.6}
\usepackage{mathrsfs}
\usepackage{times}
\usepackage[colorlinks,
citecolor=blue,linkcolor=blue,urlcolor=blue]{hyperref}
\usepackage{scalerel}
\usepackage{tensor}
\usepackage{etoolbox}

\makeatletter
\def\@mkboth#1#2{}
\newlength\appendixwidth
\preto\appendix{\addtocontents{toc}{\protect\patchl@section}}
\newcommand{\patchl@section}{%
  \settowidth{\appendixwidth}{\textbf{Appendix }}%
  \addtolength{\appendixwidth}{1.5em}%
  \patchcmd{\l@section}{1.5em}{\appendixwidth}{}{\ddt}%
}
\makeatother

\def\eqref#1{(\ref{#1})}

\usepackage{cancel}

\newcommand{\be}{\begin{equation}}
\newcommand{\ee}{\end{equation}}
\newcommand{\bea}{\begin{eqnarray}}
\newcommand{\eea}{\end{eqnarray}}

\newcommand\reallywidehat[1]{\arraycolsep=0pt\relax%
\begin{array}{c}
\stretchto{
  \scaleto{
    \scalerel*[\widthof{\ensuremath{#1}}]{\kern-.5pt\bigwedge\kern-.5pt}
    {\rule[-\textheight/2]{1ex}{\textheight}} 
  }{\textheight} %
}{0.5ex}\\           
#1\\                 
\rule{-1ex}{0ex}
\end{array}
}

\begin{document}

\title[]{Non-analytic behavior of the Loschmidt echo in XXZ spin chains: exact results}
\author{Lorenzo Piroli$^{1}$, Bal\'{a}zs Pozsgay$^{2,3}$, Eric Vernier$^{4}$}
\address{$ˆ1$ SISSA and INFN, via Bonomea 265, 34136 Trieste, Italy}
\address{$^2$ Department of Theoretical Physics, Budapest University
	of Technology and Economics, 1111 Budapest, Budafoki \'{u}t 8, Hungary}
\address{$ˆ3$ BME Statistical Field Theory Research Group, Institute of Physics,
	Budapest University of Technology and Economics, H-1111 Budapest, Hungary}
\address{$^4$\,The Rudolf Peierls Centre for Theoretical Physics, Oxford University, Oxford, OX1 3NP, United Kingdom.}

\ead{lpiroli@sissa.it, pozsgay.balazs@gmail.com, eric.vernier@physics.ox.ac.uk}
\date{\today}

\begin{abstract}
We address the computation of the Loschmidt echo in interacting
integrable spin chains after a quantum quench. We focus on the
massless regime of the XXZ spin-1/2 chain and present exact results
for the dynamical free energy (Loschmidt echo per site) for a special
class of integrable initial states. For the first time we are able to
observe and describe points of non-analyticities using exact methods,
by following the Loschmidt echo up to large real times. The dynamical
free energy is computed as the leading eigenvalue of an appropriate
Quantum Transfer Matrix, and the non-analyticities arise from the level crossings of this matrix. Our exact results are
expressed in terms of ``excited-state'' thermodynamic Bethe ansatz
equations, whose solutions involve non-trivial Riemann surfaces. By evaluating our formulas, we provide explicit numerical results for the quench from the N\'eel state, and we determine the first few non-analytic points.
\end{abstract}

\maketitle


\section{Introduction}\label{sec:intro}

Despite their long history, in the past decade the theory of integrable models has witnessed a series of unexpected developments. Among these, the most prominent one is arguably the realization that analytic techniques from integrability, traditionally tailored for ground-state and thermal physics, provide powerful tools also out of equilibrium \cite{CaEM16}.

A simple but physically interesting protocol which has proven to be within the reach of integrability has been the one of quantum quenches \cite{cc-06}: a system is prepared in some well-defined state $|\Psi_0\rangle$ and left to evolve unitarily with some Hamiltonian $H$. This problem has been  extensively studied in the past few years, since it represents an ideal and simplified setting for exploring several important questions of many-body physics out of equilibrium \cite{PSSV11,ViRi16,EsFa16,BeDo16Review,VMreview}. Among these, the problem of relaxation has represented a main motivation in the study of quantum quenches \cite{RiDO08}: based on the knowledge of the initial state $|\Psi_0\rangle$, can we predict the local stationary properties of the system at large times?

A complete answer to this question has been obtained: while in the generic case the post-quench stationary properties are thermal \cite{RiDO08}, in integrable systems they are locally captured by a Generalized Gibbs Ensemble (GGE) \cite{RDYO07,CaEF11,FaEs13,Pozs13GGE,IDWC15,IlQC17,PoVW17,PiVC16}. The latter is analogous to the Gibbs statistical ensemble, but it is constructed by taking into account, in addition to the Hamiltonian, all higher local conserved operators \cite{IMPZ16}. Besides this established conceptual picture, recent research has also provided us with quantitative means to make explicit predictions in concrete cases; a relevant example is the Quench Action approach \cite{CaEs13,Caux16}, which allows us to compute, in several cases of interest, the stationary values of local correlations at large times \cite{BeSE14,DWBC14,WDBF14,PMWK14,BePC16,Bucc16,PiCE16,AlCa16_QA,BeTC17}.

A problem which has turned out to be much harder, from the analytical point of view, is the computation of the \emph{full} real-time dynamics of local observables. Indeed, most of the work in this direction has been limited to the analysis of free systems \cite{CaEF11,cazalilla-06,bpgd-09,mc-12,se-12,csc-13,KoCC14,bkc-14,rs-14,dc-14,msca-16,bf-16,PiCa17}, while only a few studies exist in the interacting case \cite{ia-12,mussardo-13,delfino-14,dpc-15,pe-16,cubero-16,vwed-16,kz-16,AlCa17,CuSc17,Delf17,BeSE14}, mainly employing either semi-classical \cite{kz-16,AlCa17} or field theoretical methods \cite{mussardo-13,BeSE14,delfino-14,dpc-15,pe-16,cubero-16,vwed-16,CuSc17,Delf17}. Furthermore, despite the Quench Action approach provides a formal representation for the time evolution of local observables, it is usually overwhelmingly complicated to evaluate, and so far this task was carried out only in the case of interaction quenches in one-dimensional Bose gases \cite{dpc-15}.

Partly motivated by this problem, an analytic computation of the so-called Loschmidt echo in the XXZ Heisenberg chain was initiated in \cite{Pozs13,PiPV17}. The latter is not a local quantity: it is defined as the squared absolute value of the overlap between evolved and initial states. However, it is of experimental relevance being accessible,
for example, by nuclear magnetic resonance \cite{LeUP98,PLUR00}. Most prominently, the Loschmidt echo is a central object in the study of dynamical phase transitions \cite{qslz-06,silva-08,pmgm-10,fagotti2-13,dpfz-13,hpk-13,ks-13,cwe-14,heyl-14,vths-13,as-14,deluca-14,heyl-15,kk-14,vd-14,ps-14,ssd-15,sdpd-16,zhks-16,zy-16,ps-16,Heyl17,JaJo17}, and quantum revivals \cite{sd-11,Card14,Step17,NaRa17,KrLM17}, and as such, it has received increasing attention over the past few years. Furthermore, the calculation of the Loschmidt echo represents an intermediate step towards the more ambitious goal of computing the time evolution of local observables \cite{Pozs13,PiPV17}.

A promising analytical approach to its computation was proposed in \cite{Pozs13}, which is based on the so called Quantum Transfer Matrix (QTM) formalism \cite{klum-93,suzuki-99,klumper-04}. This method can be applied quite generally to an infinite family of initial \emph{integrable} states, which have been introduced and studied in \cite{PiPV17_int}. Building upon the results of \cite{Pozs13}, a full solution to the problem of computing the Loschmidt echo for imaginary time and arbitrary initial integrable states was presented in \cite{PiPV17}, while partial results were obtained for real times. In this work we complete the programme initiated in \cite{PiPV17} and provide a full solution to the real-time problem, for which a significant amount of additional techniques has to be introduced.

Differently from \cite{Pozs13,PiPV17}, in this work we will focus on quantum quenches to the gapless regime of the XXZ Hamiltonian. The reason to do this is two-fold: on the one hand, some technical simplifications occur, which allow us to reduce the amount of unnecessary complications. On the other hand, for the particular initial states considered, the Loschmidt echo displays, in the gapless regime, non-analytic points at relatively short times \cite{as-14}; this allows us to show explicitly that our method is perfectly capable to capture them. Since we are only interested in presenting the general methods, we will focus uniquely on quenches from the N\'eel state, which has already served many times in the recent literature as a prototypical case of study \cite{bpgd-09,WDBF14,PMWK14,BrSt17}. We emphasize, however, that the method detailed in this work is general, applies for arbitrary values of the anisotropy, and can be carried out for arbitrary integrable states \cite{PiPV17_int}.

The organization of this work is as follows. In Sec.~\ref{sec:setup} we present the XXZ Hamiltonian and the quench protocol, while the QTM approach is reviewed in Sec.~\ref{sec:qtm}. We tackle the computation of the Loschmidt echo in Sec.~\ref{sec:smalltimes}, where the small-time dynamics is addressed: in this case, no additional complication arises with respect to imaginary times. The calculation of the Loschmidt echo for arbitrary time is presented in Sec.~\ref{sec:generaltimes} and Sec.~\ref{sec:full_spectrum}, where all the new analytic techniques are introduced. Explicit results for the quench from the N\'eel state are also reported and discussed. Finally, our conclusions are presented in Sec.~\ref{sec:conclusion}. The most technical part of our work is consigned to the appendix.

\section{Setup}\label{sec:setup}

\subsection{The model}

We consider the $XXZ$ spin-$1/2$ chain
\bea
H &=& \frac{J}{4} \sum_{j=1}^{L}\left[\sigma^{x}_j \sigma^{x}_{j+1}+  \sigma^{y}_j\sigma^{y}_{j+1}+ \Delta \left( \sigma^{z}_j\sigma^{z}_{j+1}-1\right)\right] \,,
\label{eq:hamiltonian}
\eea
where we take $J>0$, while $\sigma_j^{\alpha}$ are the Pauli matrices. We assume periodic boundary conditions, $\sigma_{L+1}^\alpha \equiv \sigma_1^\alpha$, and take the length $L$ to be an even integer. We indicate the associated Hilbert space as $\mathcal{H}=h_1\otimes \ldots \otimes h_L$, where $h_j\simeq \mathbb{C}^{2}$ is the local Hilbert space corresponding to the site $j$. In this work we focus on the gapless regime of the model
\be
\Delta= \cos \gamma<1\,, 
\label{eq:gamma_parameter}
\ee
with $\gamma \in \mathbb{R}$. As a technical hypothesis, we restrict to the special case of anisotropies corresponding to the so-called root of unity points, where $\gamma$ is a rational multiple of $\pi$.
We focus in particular on the simplest case
\be
\gamma = \frac{1}{p+1} \pi\,,
\label{eq:def_rational_gamma}
\ee 
where $p>1$ is an integer number.

\subsection{The quench protocol and the Loschmidt echo}

In this work we are interested in quantum quenches from a special class of integrable initial states. These have been introduced and defined in \cite{PiPV17_int} to be the states annihilated by all the local conserved operators of the Hamiltonian which are odd under space reflection. They include two-site product states and matrix product states with arbitrary bond dimensions. In order to illustrate the main ideas, we will focus on the simplest example, the well-known N\'eel state
\be
|N\rangle=|\downarrow\uparrow\ldots\downarrow\uparrow\rangle\,.
\label{eq:neel}
\ee
From our derivation, detailed in the following, it will be clear that our approach could be directly applied more generally to all the integrable states defined in \cite{PiPV17_int}. 

The Loschmidt echo is arguably the simplest quantity to compute after a quantum quench. Given the initial state $|\Psi_0\rangle$, it is defined as
\be
\mathscr{L}(t) =  \left| \langle \Psi_0 | e^{- i  H t} |  \Psi_0 \rangle  \right|^2 \,,
\label{eq:Losch}
\ee
and is a measure of the probability of finding the system close to its initial configuration. For a global quench, $\mathscr{L}(t)$ decays exponentially with the volume $L$, and it is natural to introduce the Loschmidt echo per site
\be
\ell(t) =  \left[\mathscr{L}(t)\right]^{1/L} \,,
\label{eq:losch}
\ee
or, alternatively, the return rate
\be
r(t)=\frac{1}{L}\log\mathscr{L}(t)=\log \ell(t)\,.
\label{eq:return}
\ee
For global quenches, the analytical computation of the Loschmidt echo is in general extremely hard, and most of the results in the literature are restricted to either free or conformal systems, and local quenches (see however \cite{Step17} for an analytical computation starting from a domain wall state).

An analytical approach was introduced in \cite{Pozs13,PiPV17} in the gapped regime of the XXZ Hamiltonian, where an exact calculation was presented for the partition function
\be
Z(w)=\langle \Psi_0 | e^{-w H } |  \Psi_0 \rangle\,,\qquad w\in\mathbb{C}\,.
\label{eq:partition}
\ee
In particular, the starting point of \cite{PiPV17} was considering real values of $w$ (imaginary time evolution), for which the exact solution for $Z(w)$ was obtained in terms of analytic formulas. Subsequently, analytic continuation was performed to obtain the real-time evolution of the Loschmidt echo. This procedure was shown to provide the correct result only up to a finite time $t^{\ast}$.  In this work we go beyond and present a complete derivation of the real-time Loschmidt echo, by considering complex values of the parameter $w$ from the beginning. 

In the next section, we review the Quantum Transfer matrix construction introduced in \cite{Pozs13,PiPV17}, which can be carried out straightforwardly also in the massless regime of the Hamiltonian \eqref{eq:hamiltonian}.

\section{The Quantum Transfer Matrix approach}\label{sec:qtm}

\subsection{General idea}

The idea behind the Quantum Transfer Matrix approach to the Loschmidt echo relies on interpreting it as a particular boundary
partition function; this is a natural identification which has been exploited many times in the literature \cite{silva-08,hpk-13,ks-13,cwe-14,vths-13,as-14,Card14,ps-14,vd-14,ssd-15,sdpd-16}. Here, we only briefly review the main formulas for later reference, referring in particular to \cite{PiPV17} for a detailed and pedagogical treatment.

We start by introducing the building blocks of our algebraic construction, which is based on the so-called algebraic Bethe ansatz method \cite{kbi-93}. The latter is a powerful set of techniques which allows us, among many other things, to analytically diagonalize the Hamiltonian \eqref{eq:hamiltonian}. The central object is the $R$-matrix, 
\be  
R_{12}(u) 
= 
\left(
\begin{array}{cccc}
	\sin(u+\gamma) & & & \\
	& \sin u & \sin \gamma & \\
	&\sin \gamma  & \sin u  & \\
	& & & \sin(u+\gamma)
\end{array}
\right)
\,,
\label{eq:Rmatrix}
\ee
from which, one can define the transfer matrix
\be
\tau(u) = {\rm tr}_0 \left\{R_{0,L}(u) \ldots R_{0,1}(u)\right\} \,.
\label{eq:transfermatrixt}
\ee 
By means of the identity \cite{klumper-04}
\be
\frac{\tau(-\beta/2N)\tau(-\gamma + \beta/2N )}{\sin(-\beta/2N + \gamma)^{2 L }} = 1 - \frac{2\beta}{JN\sin \gamma} H + O\left(\frac{1}{N^2}\right) \,,
\label{eq:transfer_identity}
\ee
where $\gamma$ is defined in \eqref{eq:gamma_parameter}, the subsequent action of transfer matrices can be interpreted as a discrete approximation to the unitary evolution, as it follows from the well-known Suzuki-Trotter decomposition
\be
e^{-w H }=\lim_{N \to \infty} \left(1 - \frac{w H}{N}\right)^N \,.
\label{eq:suz_trot}
\ee
Indeed, using \eqref{eq:transfer_identity}, one has 
\bea 
 \left(1 - \frac{w H}{N}\right)^N \simeq   \frac{   \left[ \tau(-\beta_w/2N) \tau(-\gamma + \beta_w/2N ) \right]^N }
  {\sin(-\beta_w/2N + \gamma)^{2 L N}}  \,,
  \label{eq:SuzukiTrotter}
\eea
where we defined
\be
\beta_w=\frac{J}{2}\sin(\gamma)w\,.
\label{eq:beta_parameter}
\ee
As we have already stressed, our approach to the computation of the partition function \eqref{eq:partition} applies to all integrable states introduced in \cite{PiPV17_int}; these constitute a large family which include matrix product states of arbitrary bond dimension. For the sake of presentation, in the following we will however restrict to two-site product states of the form
\be
|\Psi_0\rangle =|\psi_0\rangle \otimes \ldots \otimes |\psi_0\rangle\,.
\ee
Following \cite{PiPV17}, it is straightforward to simplify the partition function \eqref{eq:partition} by means of the above identities. After a few steps which are not reported here, one obtains
\bea 
\langle \Psi_0 | e^{-w H } |  \Psi_0 \rangle  = 
\lim_{N \to \infty}{\rm tr}\left[  
\mathcal{T}^{L/2}\right]\,,
\,
\label{eq:loschmidtQTM}
\eea
where we introduced the boundary quantum transfer matrix
\be
\mathcal{T}=\frac{\langle \psi_0 |T^{\rm QTM}(0)\otimes T^{\rm QTM}(0) |\psi_0 \rangle}{\left[\sin(-\beta/2N+\gamma)\right]^{4N}}\,,
\label{eq:mathcal_t}
\ee
and where $T^{\rm QTM}(u)$ is the corresponding monodromy operator
\bea
T^{\rm QTM}(u)&=&L_{2N,0}(u-\beta/2N)L_{2N-1,0}(u+\beta/2N-\gamma)\cdots\nonumber\\
 &\cdots & L_{2,0}(u-\beta/2N)L_{1,0}(u+\beta/2N-\gamma)\,.
\eea
In these formulas we have omitted the subscript $w$ in $\beta_w$. 

Eq.~\eqref{eq:loschmidtQTM} is the starting point for our derivation, along the lines of \cite{PiPV17}. Assuming that the limits of large $N$ and large $L$ can be exchanged, Eq.~\eqref{eq:loschmidtQTM} amounts to compute the $w$-dependent leading eigenvalue $\Lambda_0$ of $\mathcal{T}$; indeed we have
\bea 
 \langle \Psi_0 | e^{-w H } |  \Psi_0 \rangle  \simeq 
  \left(\lim_{N \to \infty}  
 {\Lambda_0}\right)^{L/2}
 \,.
 \label{eq:loschmidtLambda}
\eea
This problem has been completely solved in \cite{PiPV17} for arbitrary choices of the state $|\psi_0\rangle$ and $w\in \mathbb{R}$. The idea is to relate the operator $\mathcal{T}$ to an integrable transfer matrix with open boundary conditions, which can be analyzed by means of the so-called boundary algebraic Bethe ansatz~\cite{skly-88,kkmn-07,kmn-14,wycs-15}. The form of $\mathcal{T}$ explicitly depends on the initial state considered, and for generic integrable states \cite{PiPV17_int} one needs to resort to the \emph{non-diagonal} version of the latter \cite{nepomechie-02,nepomechie-04,clsw-03,fgsw-11,niccoli-12,cysw-13,nepomechie-13,NepoWang14}, as explicitly worked out in \cite{PiPV17}.
In this work, to avoid unnecessary complications, we will restrict to initial states which only require us to deal with the simpler \emph{diagonal} boundary algebraic Bethe ansatz. Furthermore, as repeatedly stressed, we will work in the gapless regime of the Hamiltonian \eqref{eq:hamiltonian}, contrary to \cite{PiPV17}. In the next section we will thus review the technical aspects of the boundary algebraic Bethe ansatz in the gapless case, referring the reader to the literature for a more comprehensive treatment~\cite{skly-88,kkmn-07,kmn-14,wycs-15}.

\subsection{The boundary algebraic Bethe ansatz}

The central object of the boundary algebraic Bethe ansatz approach is the boundary transfer matrix
 \be
\tau_B(u) = {\rm tr}_0 \{ K^+(u) T_1(u) K^{-}(u) T_2(u) \} \,,
\label{eq:boundary_tf}
\ee
where
\bea
T_1(u)&=&L_{2N}(u)\ldots L_1(u)\,,\\
L_j(u)&=&R_{0,j}(u-\xi_j)\,,\\
T_2(u)&=&R_{1,0}(u+\xi_1-\gamma)\ldots R_{2N,0}(u+\xi_{2N}-\gamma)\,.
\eea
Here $R_{ij}(u)$ is the $R$-matrix introduced in \eqref{eq:Rmatrix}.
The inhomogeneities $\xi_j$ are parameters which can be chosen arbitrarily, and the trace in \eqref{eq:boundary_tf} is performed over the auxiliary space $h_0\simeq \mathbb{C}^{2}$. The $2\times 2$ boundary matrices $K^{\pm}(u)$ have to be chosen to satisfy appropriate non-linear relations known as reflection equations \cite{skly-88}. In the diagonal case of interest in this work, the general solution to the latter reads
\bea
K^{\pm}(u)&=&K(u\pm \gamma/2,\xi_{\pm})\,\\
K(u,\xi)&=&
\left(\begin{array}{cc}
	\sin\left(\xi+u\right)& 0 \nonumber\\
	0&\sin(\xi-u)
\end{array}\right)
\label{eq:k_matrix_diag}\,.
\eea

To simplify the discussion, we restrict from the beginning to quantum quenches from the N\'eel state \eqref{eq:neel}. On the one hand, it has already served many times in the recent literature as a prototypical case of study \cite{bpgd-09,WDBF14,PMWK14,BrSt17}; on the other hand, it is straightforward to apply the techniques introduced in the following to treat more general integrable states, so that this is by no means a restrictive choice. Specifying the initial state to be the N\'eel state amounts to choosing $|\psi_0\rangle=|\uparrow\downarrow\rangle$. Then, following \cite{PiPV17}, one can show that in this case one has
\bea 
\mathcal{T}=  - \frac{  1 }
  {\sin(-\beta/2N + \gamma)^{4 N}}  
  \frac{1}{\sin^2\gamma }
 \tau_B(0) \,,
 \label{eq:BQTMdef}
 \eea 
provided that the inhomogeneities and boundary parameters in $\tau_{B}(u)$ are chosen as 
\bea
\xi_{2j+1}&=&\beta/2N\,,\label{eq:inh_1}\\
\xi_{2j}&=&\gamma-\beta/2N\,,\label{eq:inh_2}
\eea
and 
\bea
\xi_{\pm}&=&\mp \frac{\gamma}{2}\,.
\label{eq:xipm}
\eea
Eq.~\eqref{eq:BQTMdef} is a key relation, which allows us to resort to integrability based methods to compute the Loschmidt echo. Indeed, the problem of computing the partition function \eqref{eq:partition} is reduced to finding the leading eigenvalue of the boundary transfer matrix \eqref{eq:boundary_tf}, where the dependence on $w$ is encoded in the inhomogeneities \eqref{eq:inh_1} and \eqref{eq:inh_2}.

The diagonalization of the boundary transfer matrix \eqref{eq:boundary_tf} can be performed analytically within the framework of the boundary algebraic Bethe ansatz~\cite{skly-88,kkmn-07,kmn-14,wycs-15}. Note that for diagonal boundaries, the transfer matrix \eqref{eq:boundary_tf} commutes with the operator counting the number $R$ of down spins. Then, the eigenstates of the open transfer matrix \eqref{eq:boundary_tf} are constructed in terms of a set of complex numbers $\{\lambda_j\}_{j=1}^{R}$, which are the so-called Bethe roots or rapidities. They are obtained as the solution of a non-linear set of equations (Bethe equations) \cite{kkmn-07}, which in our case read
\bea
\fl\left[ 
\frac{\sin(\lambda_j + \beta/2N -\gamma ) \sin(\lambda_j - \beta/2N)}{\sin(\lambda_j - \beta/2N + \gamma ) \sin(\lambda_j + \beta/2N)}
\right]^{2N} 
\prod_{k\neq j}^R \frac{\sin(\lambda_j - \lambda_k + \gamma)\sin(\lambda_j + \lambda_k + \gamma)}{\sin(\lambda_j - \lambda_k - \gamma)\sin(\lambda_j + \lambda_k - \gamma)}
\nonumber \\
\times \frac{\sin(\lambda_j - (\xi_+- \gamma/2))\sin(\lambda_j - (\xi_- - \gamma/2))}{\sin(\lambda_j +(\xi_+- \gamma/2))\sin(\lambda_j + (\xi_- - \gamma/2))}
=
1\,.
\label{eq:bethe_eq_ex}
\eea
Each set of rapidities $\bm{\lambda} \equiv  \{\lambda_j\}_{j=1}^R$ associated with the different eigenstates uniquely specifies the corresponding eigenvalue $\tau_{\bm{\lambda}}(u)$ of the boundary transfer matrix $\tau_{B}(u)$ in \eqref{eq:boundary_tf}. First, given a set $\bm{\lambda} \equiv  \{\lambda_j\}_{j=1}^R$, we introduce the doubled set
\be
\{\tilde{\lambda}_k\}_{k=1}^{2R}=\{\lambda_k\}_{k=1}^{R}\cup\{-\lambda_k\}_{k=1}^{R}\,,
\label{eq:doubledroots}
\ee
so that the corresponding eigenvalue reads \cite{kkmn-07}
\be
\tau_{\bm{\lambda}}(u)= \omega_1(u)\phi(u+\gamma/2)\frac{Q(u-\gamma)}{Q(u)}+\omega_2(u)\phi(u-\gamma/2)\frac{Q(u+\gamma)}{Q(u)}\,,
\label{eq:TQrelation_diagonal}
\ee
where we have defined
\bea
Q(u)& \equiv &\prod_{k=1}^{2R}\sin(u-\tilde{\lambda}_k)\,,\\
\phi(u) & \equiv & \prod_{k=1}^{2N}\sin\left(u-\gamma/2+\xi_k\right)\sin\left(u+\gamma/2-\xi_k\right)\,,\label{eq:phi_function}\\
\omega_1(u)&=&\frac{\sin(2u+\gamma)\sin(u+\xi^+-\gamma/2)\sin(u+\xi^--\gamma/2)}{\sin(2u)}\,,\label{eq:omega1_function}\\
\omega_2(u)&=&\frac{\sin(2u-\gamma)\sin(u-\xi^++\gamma/2)\sin(u-\xi^-+\gamma/2)}{\sin(2u)}\,.\label{eq:omega2_function}
\eea
Eq.~\eqref{eq:TQrelation_diagonal} is sometimes referred to as the $T-Q$ relation \cite{wycs-15}. We note that the Bethe equations \eqref{eq:bethe_eq_ex} can be rewritten in terms of the functions introduced above as  
\be
\frac{\omega_2(\lambda_j)}{\omega_1(\lambda_j)}\frac{Q(\lambda_j+\gamma)\phi(\lambda_j-\gamma/2)}{Q(\lambda_j-\gamma)\phi(\lambda_j+\gamma/2)}=-1\,.
\label{eq:new_bethe_eq_diagonal}
\ee
Eq.~\eqref{eq:TQrelation_diagonal} provides a formal solution to the problem of diagonalizing the transfer matrix~\eqref{eq:transfermatrixt} for finite $N$.

From Eqs.~\eqref{eq:loschmidtLambda} and \eqref{eq:BQTMdef} we see that our goal is to compute the leading eigenvalue in the limit $N\to\infty$. Two main difficulties, accordingly, arise: the first one consists in the determination of the Bethe roots $\bm{\lambda} \equiv  \{\lambda_j\}_{j=1}^R$ corresponding to the leading eigenvalue at finite $N$; the second pertains the computation of the limit $N\to\infty$ of the expression \eqref{eq:TQrelation_diagonal}.

The configuration of Bethe roots depends on $w$ [defined in \eqref{eq:loschmidtLambda}]. For each ``time'' $w$, eigenvalues which are close to each other might correspond to very different sets of rapidities; as $w$ varies each set of Bethe roots also varies continuously. However, it might happen that two eigenvalues undergo a \emph{crossing}: accordingly, the set of Bethe roots corresponding to the leading eigenvalue might change abruptly as $w$ varies smoothly, which makes the computation of the Loschmidt echo non-trivial. It turns out that for $w\in\mathbb{R}$ no crossing occurs, and the Bethe roots associated with the leading eigenvalue have a similar qualitative behavior for all values of $w\in \mathbb{R}$ \cite{Pozs13}. This is not the case for imaginary times $w=it$ ($t\in \mathbb{R}$), considered in this work. In order to set the stage, we will start in Sec.~\ref{sec:smalltimes} by treating the case of small times $t$, where no crossing arises. This allows us to focus on a single eigenvalue, for which the configuration of Bethe roots is relatively simple. In Sec.~\ref{sec:generaltimes} the same eigenvalue is computed for arbitrary values of $t$, for which the technical treatment becomes necessarily more sophisticated. Finally, for large times crossings arise, as it will be discussed in Sec.~\ref{sec:full_spectrum}: in this case, our strategy will consist of computing, for each $t$, also higher eigenvalues of $\tau_{B}(u)$ and to follow their evolution continuously, keeping track of all the subsequent crossings.

\section{The Loschmidt echo at small times}
\label{sec:smalltimes}

\begin{figure}
\begin{center}
\includegraphics[scale=0.7]{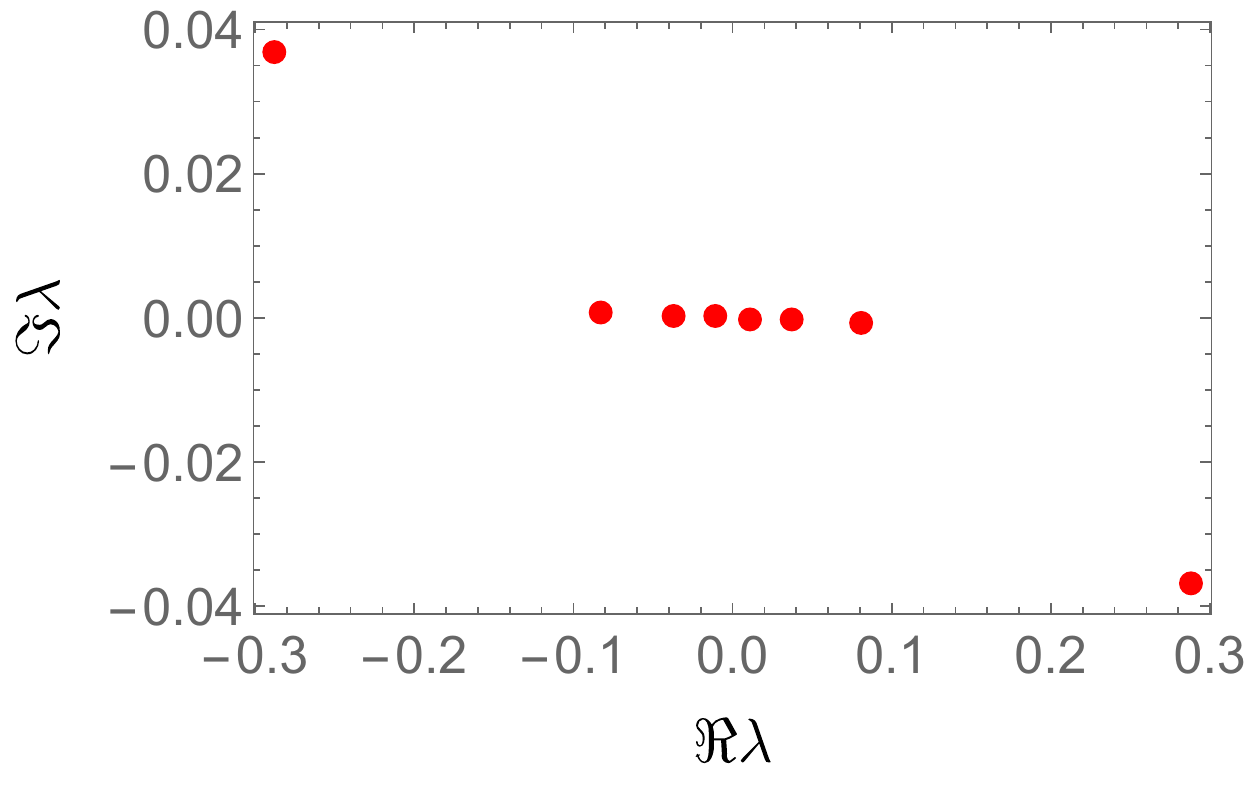}
\end{center}
\caption{
Doubled set of Bethe roots $\{ \tilde{\lambda}_i \}_{i=1}^{2N}$ associated with the leading eigenvalue of the boundary QTM in the complex-$\lambda$ plane, for $2N=8$, $\gamma=\pi/3$ and $w = it = 1 i $. The $x$- and $y$- axes correspond to real and imaginary parts of the rapidities.}
\label{fig:roots}
\end{figure}

We set $w=it$ in Eq.~\eqref{eq:partition} with $t\in\mathbb{R}$, and $t$ sufficiently small. Following \cite{PiPV17}, we start with a preliminary numerical analysis at finite values of $N$ of the eigenvalues of the boundary QTM \eqref{eq:boundary_tf}, which can be obtained by exact diagonalization. It is found that for small values of $t$ the leading eigenvalue of the boundary QTM is unique, with a finite gap with respect to the higher ones. Furthermore, it lies in the sector of zero magnetization, and therefore is associated with $R=N$ Bethe roots. For small values of $N$, these can be identified numerically following a standard procedure, by comparing the formal $T-Q$ relation \eqref{eq:TQrelation_diagonal} with explicit diagonalization of $\tau_B(u)$, as already done in \cite{PiPV17}. An example of a configuration of Bethe roots for the leading eigenvalue of $\tau_{B}(u)$ is displayed in Fig.~\ref{fig:roots}.

The Bethe roots do not arrange, in the limit $N\to\infty$, according to a smooth rapidity distribution function. In order to take the infinite-$N$ limit, then, two routes can be followed. The first one consists in writing down a single nonlinear integral equation for an appropriate auxiliary function, as done in \cite{Pozs13}. While this can be easily done for the N\'eel state, serious complications arise in the study of more general integrable states, corresponding to nondiagonal boundary conditions \cite{PiPV17}. Accordingly, in order to keep the discussion as general as possible, we will follow the second approach, detailed in \cite{PiPV17}, which is based on the so-called $Y$-system relations\cite{kns-11}, and for which no additional complication arises in the case of generic states. In the following subsections we will introduce the $Y$-system and review how the latter can be exploited to obtain directly the spectrum of the boundary QTM.

\subsection{The $Y$-system}
\label{sec:Ysystem}

It is an established result that the boundary transfer matrix $\tau_B(u)$ can be used to build an infinite family of transfer matrices $\{t_j(u)\}_{j=0}^{\infty}$ via the so-called fusion procedure \cite{zhou-95,mn-92,zhou-96}, in complete analogy with the well-known case of periodic boundary conditions \cite{kns-11}. In the following, we outline the aspects of this construction which are relevant for our work.

The fused transfer matrices $t_{j}(u)$ act on the same space as the transfer matrix $\tau_B(u)$ and form a commuting set of operators, namely
\be
\left[t_j(u),t_k(w)\right]=0\,,\qquad j,k=0,1,\ldots\,.
\label{eq:commutation}
\ee
They can be obtained recursively as 
\bea
t_0(u) &=& \mathbf{1}\,,  \nonumber\\
t_1(u) &=& \tau_B(u)\,,  \nonumber\\
t_j(u) &=& t_{j-1}\left(u-\frac{\gamma}{2}\right)t_{1}\left(u+(j-1)\frac{\gamma}{2}\right)\nonumber\\
  &-& f\left(u + (j-3)\frac{\gamma}{2}\right) t_{j-2}(u-\gamma) \,, \qquad j\geq 2 \,.
\label{eq:Tsys1}
\eea
Here $f(u)$ is defined as 
\be
f(u - \gamma/2) =  \phi(u+\gamma) \phi(u-\gamma) \omega_1 (u+\gamma/2) \omega_2 (u-\gamma/2)  \,,
\label{eq:fdiagonal}
\ee
where the functions  $\phi(u)$, $\omega_1(u)$ and $\omega_2(u)$, are defined in \eqref{eq:phi_function}, \eqref{eq:omega1_function} and \eqref{eq:omega2_function} respectively. Let us point out that the results presented in this paragraph hold for generic values of the inhomogeneities $\xi_i$ and of the boundary spectral parameters $\xi_\pm$ (note that they can also be extended to the case of non-diagonal boundary reflection matrices \cite{PiPV17}). Finally, from Eqs.~\eqref{eq:Tsys1} one can derive the relation
\be 
t_j\left( u+\frac{\gamma}{2} \right)t_j\left( u-\frac{\gamma}{2} \right) 
= 
t_{j+1}( u ) t_{j-1}( u )  + \Phi_j(u) \,, \qquad j\geq 1\,,
\label{eq:t_system}
\ee
where 
\be 
\Phi_j(u) = \prod_{k=1}^j f\left[ u - (j+2-2k)\frac{\gamma}{2} \right]\,. 
\ee 

The set of relations \eqref{eq:t_system} is usually referred to as $T$-system. From the latter one can derive a new set of functional relations, the so-called $Y$-system, which is expressed in terms of the operators
\be
Y_j(u)=\frac{t_{j-1}(u)t_{j+1}(u)}{\Phi_j(u)}\,, \qquad j\geq 1\,,
\label{eq:y_function}
\ee
with the choice $Y_0\equiv 0$. From this definition, and the $T$-system \eqref{eq:t_system}, the following $Y$-system is readily derived
\be
Y_j\left(u+\frac{\gamma}{2}\right)Y_j\left(u-\frac{\gamma}{2}\right)=\left[1+Y_{j+1}\left(u\right)\right]\left[1+Y_{j-1}\left(u\right)\right]\,.
\label{eq:y_system}
\ee

\subsection{Truncation of the $Y-$ system at the root of unity}

In general, the $Y$-system \eqref{eq:y_system} consists of an infinite number of functional relations. This is not an issue, as for practical purposes of numerical evaluation of the Loschmidt echo it can be truncated to a finite number $n_{\rm MAX}$ of them, introducing an error which decreases rapidly as $n_{\rm MAX}$ is increased \cite{PiPV17}. There exists, however, a particular case where an exact truncation takes place, and the infinite system is exactly equivalent to a finite one: namely when the parameter $q= e^{i \gamma}$ is a root of unity. In this work we will restrict to this case, in order to reduce the number of unnecessary complications. As an additional simplification, we will impose another restriction to the values of $\gamma$, which we will choose to be of the form \eqref{eq:def_rational_gamma}, with $p>1$ integer. This makes the final form of the $Y$-system particularly simple. Generalization to the case $\gamma=q\pi/p$, with $q,p>1$ integers is possible, but will not be discussed here.

For the values of $\gamma$ in Eq.~\eqref{eq:def_rational_gamma}, an exact truncation of the $Y$-system takes place due to an additional relation between the fused transfer matrices $t_{p+1}$ and $t_{p-1}$, which can be traced back to the representation theory of the underlying quantum group $U_q(sl_2)$ with $q=e^{i \gamma}$. Such a relation was originally observed for the periodic chain in \cite{Kuniba} (see also \cite{BeGF00,BeSt99}), and for general integrable open boundaries in \cite{nepomechie-02,Nepo03}. Recasting the results of the latter in our notations, for the particular case of diagonal boundary reflection matrices, we find
\be 
t_{p+1}(u)  = \frac{\Phi_p(u)}{\Psi(u)^2} t_{p-1}(u) + 2 \cosh(\alpha(u)) \frac{\Phi_p(u)}{\Psi(u)}  \mathbf{1} \,,
\label{eq:truncationT}
\ee 
where we have defined
\bea
{\Psi}(u) &=&\tilde{\Psi}(u)  g(u) \\
 \tilde{\Psi}(u) &=&  (-1)^{2N+1} \frac{\sin(2u)}{\sin(2u+2\gamma)}\phi\left( u-\frac{\pi}{2} \right)\nonumber\\
&\times & \prod_{j=1}^{p-1} \frac{\sin(2u+2j \gamma)}{\sin(2u + (2j+1)\gamma)}
\prod_{j=2}^{p-1} \phi\left( u+j\gamma-\frac{\pi}{2} \right)\\
 g(u) &=&
 \prod_{\xi = \xi_+, \xi_-}
\frac{  
\sin\left[(p+1)\left( u + \xi + \frac{\pi}{2} \right) \right]^{1/2}}{
2^{p}\cos(u+ \xi)
}\nonumber\\
&\times &\frac{\sin\left[(p+1)\left( u - \xi + \frac{\pi}{2} \right) \right]^{1/2}
}{\cos(u- \xi)}
\,,\\
 \alpha(u)  &=&    
\frac{1}{2}\ln \left(\frac{  \sin\left[(p+1)\left( u + \xi_+ + \frac{\pi}{2} \right) \right] \sin\left[(p+1)\left( u + \xi_- + \frac{\pi}{2} \right) \right]}{ \sin\left[(p+1)\left( u - \xi_+ + \frac{\pi}{2} \right) \right] \sin\left[(p+1)\left( u - \xi_- + \frac{\pi}{2} \right) \right]}  \right)
\hspace{-0.15cm}\,. 
\eea
Noticing now that 
 \be \Phi_{p-1}(u)   = {\Psi}\left( u + \frac{\gamma}{2} \right) {\Psi}\left( u  - \frac{\gamma}{2} \right)   \,,
\ee
making use of \eqref{eq:truncationT} and of the definitions \eqref{eq:y_function}, the following truncated $Y-$system can be obtained
\bea
\hspace{-1cm} Y_j\left(u+\frac{\gamma}{2}\right) Y_j\left(u-\frac{\gamma}{2}\right)   = \left[1+Y_{j+1}(u) \right] \left[1+Y_{j-1}(u) \right] \,, \quad j=1,\ldots p-1\,, \\
\hspace{-1cm} 1 + Y_{p}(u) = 1 +  2\cosh[\alpha(u)] K(u) +  K(u)^2\,,  \label{closure2}  \\
\hspace{-1cm} 1+ Y_{p-1}(\alpha) =    K\left(u+\frac{\gamma}{2}\right)K\left(u-\frac{\gamma}{2}\right) \,,
\eea
where
\be 
K(u) =  \frac{t_{p-1}(u)}{\Psi(u)} \,.
\ee  

The $Y$-system can be further simplified once we impose the boundary parameters to take the values of interest in the present problem, namely $\xi_\pm = \mp \gamma/2$, cf. Eq.~\eqref{eq:xipm}. Indeed, in this case one has $\alpha(u) =0$, and consequently we obtain 
\bea
\hspace{-1cm} Y_j\left(u+\frac{\gamma}{2}\right) Y_j\left(u-\frac{\gamma}{2}\right)   = \left[1+Y_{j+1}(u) \right]\left[1+Y_{j-1}(u) \right] \,,\quad  \ j=1,\ldots ,p-1\,, \label{eq:recursion_eigen} \\
\hspace{-1cm} 1 + Y_{p}(u) =  \left[ 1 +  K(u) \right]^2  \\
\hspace{-1cm} 1+ Y_{p-1}(u) =  K\left(u+\frac{\gamma}{2}\right)K\left(u-\frac{\gamma}{2}\right) \,.
\label{eq:Ysystemfinal}
\eea
We note that this form of the $Y$-system, which is particularly convenient from the computational point of view, holds whenever $\xi_+ = \xi_-$ or $\xi_+ = - \xi_-$. 

\subsection{From the $Y$-system to the Loschmidt echo}
\label{sec:from_y_to_LE}

The importance of the above construction for the computation of the Loschmidt echo is that it allows us to express the latter in terms of the solution of a system of non-linear equations; in turn, these can be easily evaluated numerically to yield the exact value in the infinite-$N$ limit. The same idea was already exploited in \cite{PiPV17}, and is reviewed in the following.

First, from Eq.~\eqref{eq:commutation}, we see that the operators $Y_j$ commute with one another, with the transfer matrices, and by construction with the global magnetization $S_z$ (since we are restricting to diagonal boundary conditions). Accordingly, the set of functional relations \eqref{eq:y_system} can be understood at the level of individual eigenvalues of the operators $Y_j(u)$. In the following, we indicate as $y_j(\lambda)$ the eigenvalue of $Y_{j}(i\lambda)$ (note that a rotation of $\pi/2$ in the complex plane of the argument of $y_j(\lambda)$ has been performed for convenience); we will refer to $y_{j}(\lambda)$ as the $Y$-functions. They satisfy the $Y$-system
\bea
\hspace{-1cm} y_j\left(\lambda+i\frac{\gamma}{2}\right) y_j\left(\lambda-i\frac{\gamma}{2}\right)   = [1+y_{j+1}(\lambda) ] [1+y_{j-1}(\lambda) ] \,, \quad j=1,\ldots p-1\,,  \label{eq:Ysystemlambda1}\\
\hspace{-1cm} 1 + y_{p}(\lambda) =  \left[ 1 +  \kappa(\lambda) \right]^2 \,,  \label{eq:Ysystemlambda2}\\
\hspace{-1cm} 1+ y_{p-1}(\lambda) =  \kappa\left(\lambda+i\frac{\gamma}{2}\right)\kappa\left(\lambda-i\frac{\gamma}{2}\right) \,,
\label{eq:Ysystemlambda3}
\eea
where $\kappa(\lambda)$ denotes the eigenvalue of the operator $K(i\lambda)$.
It will be useful to know the asymptotic behavior of the $Y$-functions on Bethe states as $\lambda\to \pm  \infty$. First, we make use of the simple relation for the magnetization $|S^z|$ of a given eigenstate in terms of the number $R$ of the corresponding Bethe roots,
\be 
|S^z| = N - R  \,.
\label{eq:relationRSz}
\ee 
Then, from the $T-Q$ relation \eqref{eq:TQrelation_diagonal}, we can deduce 
\be 
1 + y_1( \pm  \infty)  \sim 4 \cos^2(2 \gamma S_z) \,.
\label{eq:Y1limit}
\ee 
Finally, the asymptotic behavior of the higher $Y$-functions $y_j$, $j\geq 2$, is easily obtained by recursion using \eqref{eq:recursion_eigen}.

Following \cite{PiPV17}, we define the normalized boundary transfer matrix  
\be
\mathcal{T}(\lambda)  = -\frac{1}{\left( \sinh\left( \lambda - i \frac{\beta}{2N} + i \gamma \right) \sinh\left( -\lambda - i \frac{\beta}{2N} + i \gamma \right) \right)^{2N}}  \frac{\tau_{\rm B}(i\lambda)}{ \mathcal{N}(\lambda)}\,,
\label{eq:Lambdadef}
\ee
where 
\be
\mathcal{N}(\lambda) = -\sinh(\lambda+i \gamma)\sinh(\lambda-i \gamma) - \sinh(\lambda)^2   \,,
\ee
so that $\mathcal{T}(0)$ coincides with the operator $\mathcal{T}$ in Eq.~\eqref{eq:BQTMdef}. For each time $t$, we indicate with $\{\Lambda^{t}_{\ell}(\lambda)\}_{\ell=0}^{\infty}$ the set of eigenvalues of the corresponding operator $\mathcal{T}(\lambda)$ [the dependence on $t$ is through the parameter $\beta$, defined in \eqref{eq:beta_parameter}]. With each eigenvalue $\Lambda^{t}_{\ell}(\lambda)$, is associated a set of $Y$-functions $\{y^{(\ell)}_{j}(\lambda)\}_{j=1}^{p}$, and the following relations can be derived \cite{PiPV17} 
\bea
\hspace{-1cm}1 + y^{(\ell)}_1(\lambda ) = (1+ \tilde{y}_1(\lambda)) \left(  \frac{\sinh\left(  \lambda + i \left( \frac{\beta}{2N} - \frac{\gamma}{2} \right) \right)\sinh\left(  \lambda - i \left( \frac{\beta}{2N} - \frac{\gamma}{2} \right) \right)}{\sinh\left(  \lambda + i \left( \frac{\beta}{2N} + \frac{\gamma}{2} \right) \right)\sinh\left(  \lambda - i \left( \frac{\beta}{2N} + \frac{\gamma}{2} \right) \right)} \right)^{2N} \nonumber\\
\times \Lambda^t_\ell\left( \lambda + i \frac{\gamma}{2} \right)
\Lambda^t_\ell\left( \lambda - i \frac{\gamma}{2} \right) \,,
\label{eq:LambdaY1}
\eea
where 
\bea 
1 + \tilde{y}_1(\lambda)  &=& \frac{ \mathcal{N}\left( \lambda + i \frac{\gamma}{2} \right)
	\mathcal{N}\left( \lambda - i \frac{\gamma}{2} \right)  }{ \chi(\lambda)} \,, \nonumber\\  
\chi(\lambda) &=& 
\frac{\sinh(2\lambda + 2 i \gamma)\sinh(2\lambda - 2 i \gamma)}{\sinh(2\lambda +  i \gamma)\sinh(2\lambda - i \gamma)}\nonumber\\ 
&\times &\sinh\left( \lambda - i \frac{\gamma}{2} \right)  ^2\sinh\left( \lambda + i \frac{\gamma}{2} \right)  ^2\,.
\eea
Eq.~\eqref{eq:LambdaY1} is a functional relation between the eigenvalue $\Lambda_\ell(\lambda)$ and the $Y$-function $y^{(\ell)}_1(\lambda)$. In order to compute the former, and thus the Loschmidt echo \eqref{eq:loschmidtLambda}, we need a final step, namely to cast the functional relations \eqref{eq:Ysystemlambda1}-\eqref{eq:Ysystemlambda3} and \eqref{eq:LambdaY1} into a set of integral equations. This programme will be followed explicitly in the next sections.

\subsection{Short-time dynamics}

The procedure to cast the functional equations into integral ones is well-known in the literature (see e.g. \cite{FGSW11}). We summarize it here, for convenience, referring to the literature, and in particular to our previous work \cite{PiPV17}, for a detailed explanation. 

In order to write down the integral equations, one needs to take first the logarithmic derivative of both sides of Eqs.~\eqref{eq:Ysystemlambda1}-\eqref{eq:Ysystemlambda3}. Next, one performs a Fourier transform of both sides, obtaining a number of integrals along segments with non-zero imaginary parts. Finally, one moves the integration contours back to the real axis; in order to do so, one needs to take into account all the singularities of the logarithmic derivatives, which correspond to the zeros and poles of the functions $y_j(\lambda)$ and $\kappa(\lambda)$ inside the so called {\it physical strip}. This is the region of the complex-$\lambda$ plane defined by
\be
|\Im \lambda | \leq \frac{\gamma}{2} \,, \qquad -\infty < \Re \lambda < \infty  \,.
\label{eq:physical_strip}
\ee
After performing these steps, one is left with a set of equations of the form 
\be
\reallywidehat{\log y_j} = \frac{1}{2\cosh k \gamma}  \left[
\reallywidehat{\log\left(1+y_{j+1}\right)} 
+
\reallywidehat{\log\left(1+y_{j-1}\right)}  
\right]
+ 
\ldots \,, 
\label{eq:fourier_equations}
\ee
where the $\ldots$ denotes additional contributions coming from the poles and zeros of the functions $y_j(\lambda)$ and $\kappa(\lambda)$. Here, the following notation for the Fourier transform has been used
\bea
\hat{f}(k) &=& \int_{-\infty}^{\infty} d\lambda\, e^{ i k \lambda} f(\lambda) \,, \qquad k\in \mathbb{R} \,,\label{eq:fourier_1}
\eea
so that its inverse reads
\bea
f(\lambda) &=& \frac{1}{2\pi}\int_{-\infty}^{\infty} dk\, e^{- i k \lambda} \hat{f}(k) \,, \qquad \lambda \in \mathbb{R} \,. \label{eq:fourier_2}
\eea
Eq.~\eqref{eq:fourier_equations} can be transformed back to real space, yielding the desired set of non-linear integral equations.

As it is clear from the above discussion, the only piece of information needed in this calculation is the location of poles and zeros of the functions $y_j(\lambda)$ and $\kappa(\lambda)$. It turns out that this can be determined analytically for small times, where no additional difficulty arises with respect to the case of imaginary-time evolution. Our analysis of the analytic structure is based on numerical inspection at finite Trotter numbers $N$ of the functions $y_j(\lambda)$ and $\kappa(\lambda)$. These can always be obtained implementing the operators $Y_j(\lambda)$ and $K(\lambda)$ for finite $N$. Numerical inspection for Trotter numbers up to $2N=8$ reveals the following analytic structure, which is found to be always present for small times $t$:
\begin{itemize}
	\item $y_1$ displays the following structure 
	\begin{itemize}
		\item $y_1$ has a zero of order $2$ at $\lambda = 0$;
		\item $y_1$ has poles  at $\lambda = \pm i \frac{\gamma}{2}$, of order $2$ for $p$ even, of order $1$ for $p$ odd;
		\item $y_1$ has poles of order $2N$ at $\lambda = \pm i \left( \frac{\gamma}{2} + \frac{\beta}{2N}  \right)$; 
		\item $y_1$ has zeros of order $2N$ at $\lambda = \pm i \left( \frac{\gamma}{2} - \frac{\beta}{2N}  \right)$;
	\end{itemize}
	\item for $j \geq 1$, the only poles or zeros in the physical strip (except possible pairs at $\pm i \gamma/2$) of $y_j$, $j\geq 2$, are a double zero at $\lambda=0$ (resp. double pole) for $j$ odd (resp. $j$ even);
	\item the only poles or zeros of $\kappa$ in the physical strip (except possible pairs at $\pm i \gamma/2$) are a double pole at $\lambda=0$ for $p$ even.
\end{itemize}
Note that additional pairs of zeros or poles of the auxiliary functions $y_j(\lambda)$ and $\kappa(\lambda)$ at $\pm i \gamma/2$ do not give contributions to the integral equations and can be neglected.

Using the above information and the procedure outlined above, we obtain easily the integral equations corresponding to the functional relations \eqref{eq:Ysystemlambda1}-\eqref{eq:Ysystemlambda3}. For $p$ odd, we have
\bea 
\fl \ln y_1 =  s\ast \ln(1+y_2 ) 
- 2 \ln \left( \coth \frac{\pi \lambda}{2 \gamma} \right) 
- 2 N \ln \left(  \frac{\cosh\left( \frac{\pi \lambda}{\gamma} \right) + \sin\left( \frac{\pi \beta}{2 N \gamma} \right)}
{\cosh\left( \frac{\pi \lambda}{\gamma} \right) - \sin\left( \frac{\pi \beta}{2 N \gamma} \right)}  \right)\,, 
\\
\fl \ln y_j =  s\ast \ln(1+y_{j-1} ) +s\ast \ln(1+y_{j+1} ) 
+(-1)^j 2 \ln \left( \coth \frac{\pi \lambda}{2 \gamma} \right) \,,
\quad  2\leq j \leq p-1\,,
\\
\fl \ln \kappa =  s\ast \ln(1+y_{p-1} ) +  2 \ln \left( \coth \frac{\pi \lambda}{2 \gamma} \right)\,.
\eea 
Here we defined
\be
s(\lambda) = \frac{1}{2\gamma \cosh\left( \frac{\pi \lambda}{\gamma}  \right)} \,,
\ee 
and introduced the convolution between two functions
\be
[f\ast g ](\lambda)=\int_{-\infty}^{\infty} d\mu \,f(\lambda-\mu)g(\mu)\,.
\ee
Analogously, for $p$ even, we obtain
\bea 
\fl \ln y_1 =  s\ast \ln(1+y_2 ) 
- 2 \ln \left( \coth \frac{\pi \lambda}{2 \gamma} \right) 
- 2 N \ln \left(  \frac{\cosh\left( \frac{\pi \lambda}{\gamma} \right) + \sin\left( \frac{\pi \beta}{2 N \gamma} \right)}
{\cosh\left( \frac{\pi \lambda}{\gamma} \right) - \sin\left( \frac{\pi \beta}{2 N \gamma} \right)}  \right)\,,
\\
\fl \ln y_j =  s\ast \ln(1+y_{j-1} ) +s\ast \ln(1+y_{j+1} ) 
+(-1)^j 2 \ln \left( \coth \frac{\pi \lambda}{2 \gamma} \right) \,,
\quad 2\leq j \leq p-1\,,\\
\fl \ln \kappa =  s\ast \ln(1+y_{p-1})\,.
\eea 
Similarly, from Eq.~\eqref{eq:LambdaY1}, one gets the following relation between $\Lambda_\ell$ and $y_1$    
\be
\ln \Lambda_\ell =  s\ast \ln\left( \frac{1+y_1}{1+\tilde{y}_1} \right)    -  s \ast \psi_N \,,
\ee
where 
\bea
\psi_N(\lambda) &=&  2 N  \ln \left(  \frac{ \cosh(2 \lambda)- \cos\left( \frac{\beta}{N} - \gamma \right)}{\cosh(2 \lambda )- \cos\left( \frac{\beta}{N} + \gamma \right)} \right)\,.
\eea

The equations above are exact at finite Trotter number $N$. It is straightforward to compute the Trotter limit using
\bea
\fl \lim_{N\to\infty}  2 N \ln \left(  \frac{\cosh\left( \frac{\pi \lambda}{\gamma} \right) + \sin\left( \frac{\pi \beta}{2 N \gamma} \right)}
{\cosh\left( \frac{\pi \lambda}{\gamma} \right) - \sin\left( \frac{\pi \beta}{2 N \gamma} \right)}  \right)= \frac{2 \pi  \beta}{\cosh(\frac{\pi\lambda}{\gamma})\gamma}=\frac{i\pi t\sin\gamma}{\cosh(\frac{\pi\lambda}{\gamma})\gamma}\,,\\
\fl \lim_{N\to\infty} 2 N  \ln \left(  \frac{ \cosh(2 \lambda)- \cos\left( \frac{\beta}{N} - \gamma \right)}{\cosh(2 \lambda )- \cos\left( \frac{\beta}{N} + \gamma \right)} \right)= \frac{4\beta \sin\gamma}{\cos\gamma-\cosh2\lambda}=\frac{2i t \sin^{2}\gamma}{\cos\gamma-\cosh2\lambda}\,,
\eea
where we set for convenience $J=1$ in \eqref{eq:beta_parameter}. The resulting equations can be solved numerically by iteration, and their validity holds until the analytical structure of the $Y$-functions remains as outlined above. Note that they are the same that one would obtain by analytic continuation of the imaginary-time result, namely for $w\in\mathbb{R}$. Indeed, it was already observed in \cite{PiPV17} that the correct real-time Loschmidt echo could be derived in this way for small times.

It was already observed in \cite{PiPV17}, however, that these equations hold only up to a given time $0<t^{\ast}<\infty$, after which they do not provide anymore the correct prediction for the Loschmidt echo. In the following, we show explicitly that this is due to the fact that at $t=t^{\ast}$ additional zeros of the $Y$-functions enter the physical strip, and new source terms of the integral equations have to be considered. This allows us to go beyond the results of \cite{PiPV17}, and compute the Loschmidt echo for intermediate and large times.

\section{Full time dependence of transfer matrix eigenvalues} 
\label{sec:generaltimes}

\begin{figure}
	\begin{tabular}{lllll}
	\includegraphics[scale=0.6]{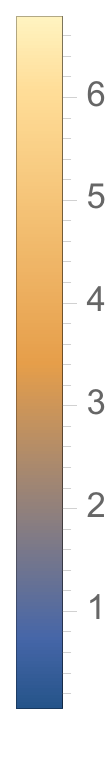} &
	\hspace{-0.35cm}\includegraphics[scale=0.55]{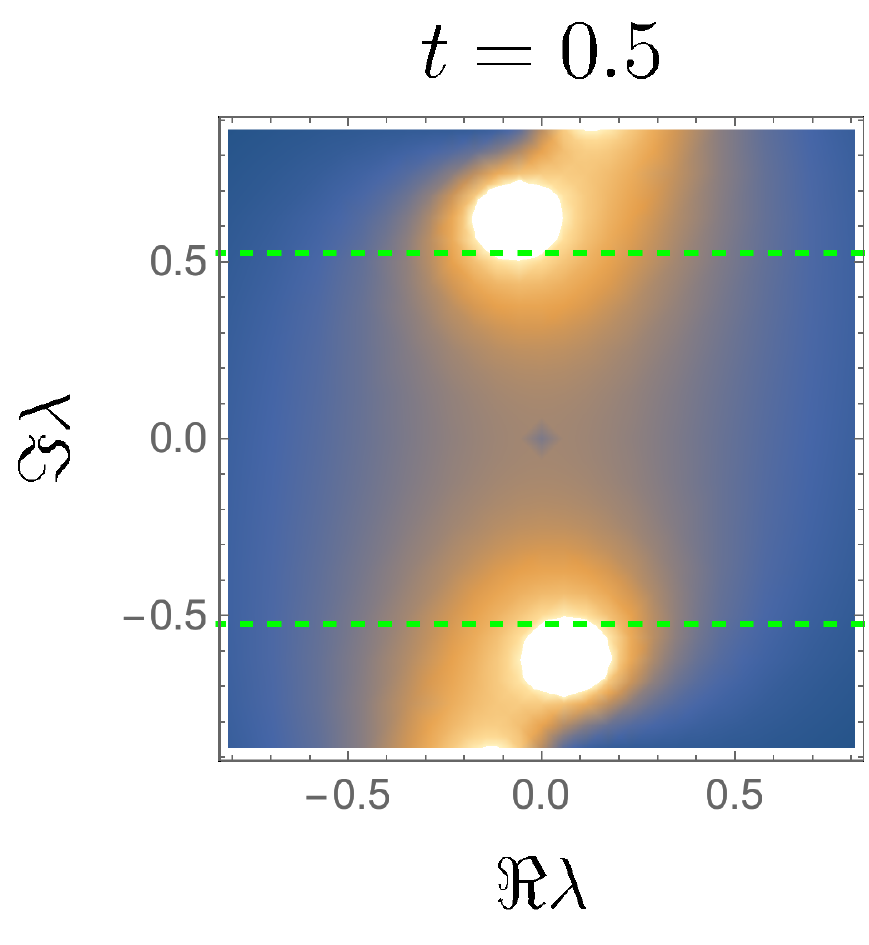} &  \hspace{-0.35cm}\includegraphics[scale=0.55]{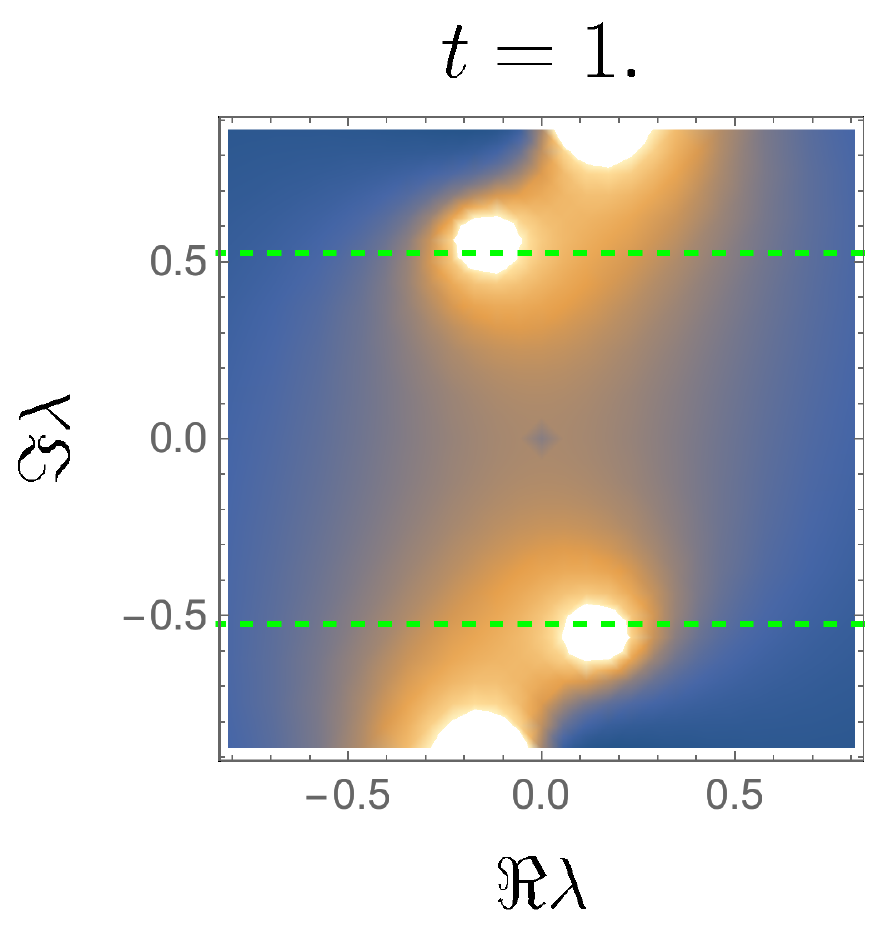}&
	\hspace{-0.35cm}\includegraphics[scale=0.55]{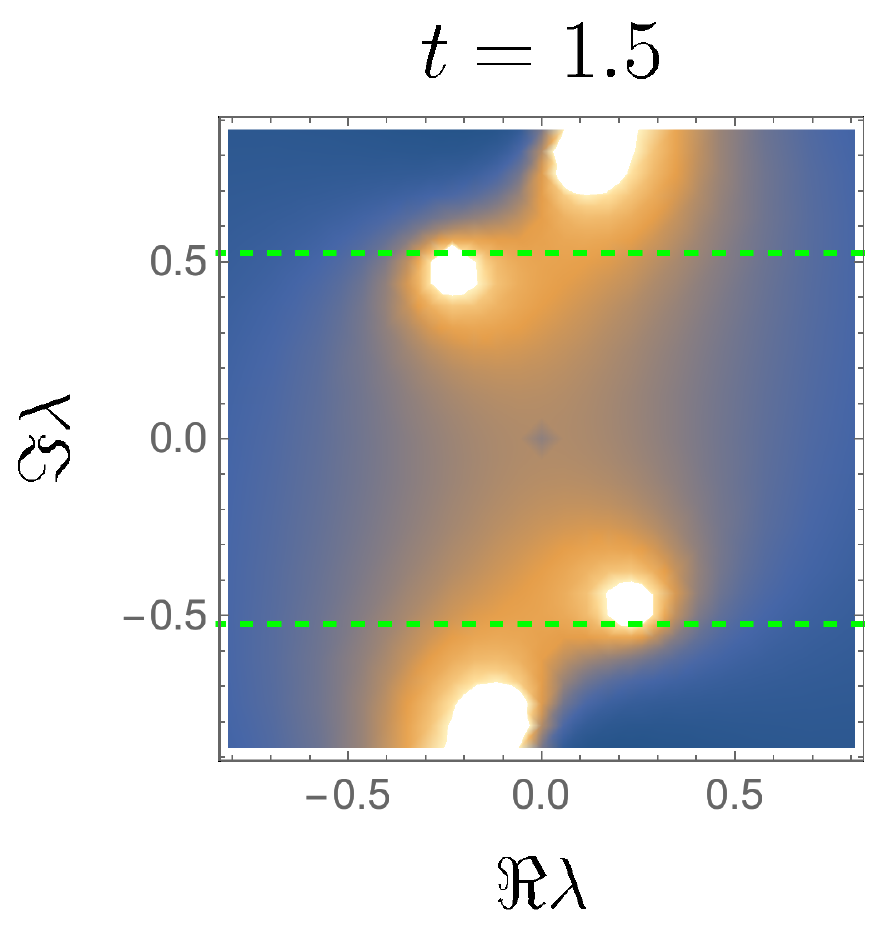}&
\end{tabular}
\caption{Density plot of $|\lambda^2  \kappa(\lambda)|^{-1}$ in the complex-$\lambda$ plane, obtained at different times from exact diagonalization at finite Trotter number 2N=6. The figure corresponds to anisotropy $\Delta=1/2$. The white zones signal additional zeros of $\kappa(\lambda)$. At a critical time $t^*$  located between $1$ and $1.5$, an additional pair of zeros enter the physical strip, whose boundaries are denoted by green dashed lines}
\label{fig:zeros}
\end{figure}

The $Y$-system encoded in Eqs.~\eqref{eq:Ysystemlambda1}-\eqref{eq:Ysystemlambda3} is valid at any time $t$. However, as we already stressed, the integral equations derived in the last section hold only up to a critical value $t=t^{\ast}$, when additional zeros and poles of the $Y$-functions enter the physical strip. This can be observed very clearly at finite Trotter number $N$ from numerical implementation of the boundary QTM, as shown in Fig.~\ref{fig:zeros}. 

Importantly, we found that the position of the additional zeros and poles for $t>t^{\ast}$ can not be determined analytically. In order to overcome this issue, we employ a procedure which was initially introduced within the framework of the so-called excited-state thermodynamic Bethe ansatz (TBA). The kinds of techniques that we will employ were first introduced in the context of thermal physics in one-dimensional solvable models \cite{KlPe93,JuKS98} and integrable quantum field theories \cite{DoTa96,BaLZ97}, and will be illustrated in the following.

For simplicity, we consider the case  $p=2$, for which the $Y$-system reads
\bea
y_1\left(\lambda+\frac{i\gamma}{2}\right)y_1\left(\lambda-\frac{i\gamma}{2}\right)&=&\left[1+\kappa(\lambda)\right]^2\,,\label{eq:y-system_p2_eq1}\\
\kappa\left(\lambda+\frac{i\gamma}{2}\right)\kappa\left(\lambda-\frac{i\gamma}{2}\right)&=&1+y_1(\lambda)\,.\label{eq:y-system_p2_eq2}
\eea
From numerical inspection, we see that no additional poles enter the physical strip, and only zeros of the $Y$-functions appear, which always come in pairs of opposite value. Suppose that additional zeros of $\kappa(\lambda)$ enter the physical strip for a given time $t$. The contributions of zeros and poles are clearly additive, so we can consider a single pair of zeros $\pm\delta^{(\kappa)}$. 
Note that we label arbitrarily one of them $\delta^{(\kappa)}$ and the other $-\delta^{(\kappa)}$. Define in the following
\be
\mathcal{I}_\delta^{(\kappa)}={\rm Im}\left[\delta^{(\kappa)}\right]\,.
\ee
Up to a global constant, applying the usual trick of integration in the complex plane we get the following term in the r.h.s. of \eqref{eq:fourier_equations}
\be
-\int_{-\infty}^{\infty}dk\,\frac{\sinh\left(k\gamma/2+{\rm sign}\left[\mathcal{I}_\delta^{(\kappa)}\right]ik \delta^{(k)}\right)}{k\cosh(k\gamma/2)} e^{- i k \lambda}\,.
\ee
This could be integrated to give
\bea
&-&\int_{-\infty}^{\infty}dk\,\frac{\sinh\left(k\gamma/2+{\rm sign}\left[\mathcal{I}_\delta^{(\kappa)}\right]ik \delta^{(k)}\right)}{k\cosh(k\gamma/2)} e^{- i k \lambda}\nonumber\\
&=&-2\pi i\, {\rm sign}\left[\mathcal{I}_\delta^{(\kappa)}\right] \left\{\mathcal{L}\left[\left(\delta^{(\kappa)}-i{\rm sign}\left[\mathcal{I}_\delta^{(\kappa)}\right]\gamma/2\right)-\lambda\right]\right.\nonumber\\
&+&\left.\mathcal{L}\left[\left(\delta^{(\kappa)}-i{\rm sign}\left[\mathcal{I}_\delta^{(\kappa)}\right]\gamma/2\right)+\lambda\right]\right\}\,,
\label{eq:intermediate_expr}
\eea
where
\be
\mathcal{L}(u)=\frac{1}{\pi}\arctan\left[\tanh\left(\frac{3\lambda}{2}\right)\right]\,.
\label{eq:ell_function}
\ee
Note that \eqref{eq:intermediate_expr} is symmetric under $\delta^{(\kappa)}\to-\delta^{(\kappa)}$, as it should.

\begin{figure}
	\begin{tabular}{lll}
		\includegraphics[scale=0.7]{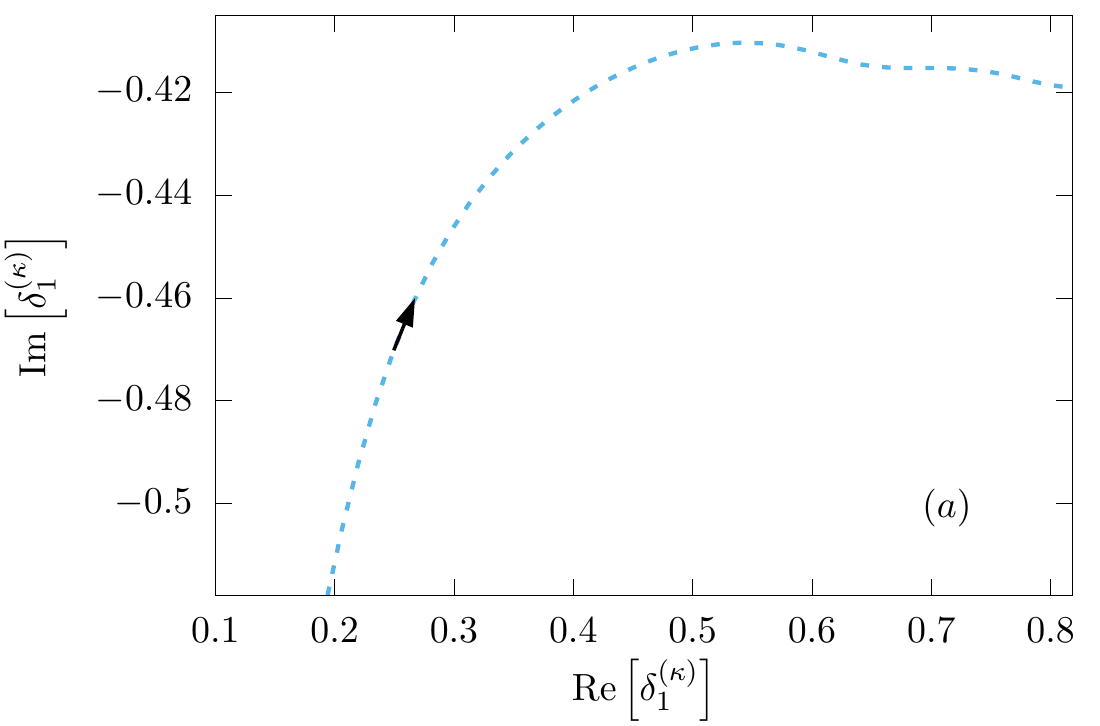} &  \hspace{-0.5cm}\includegraphics[scale=0.7]{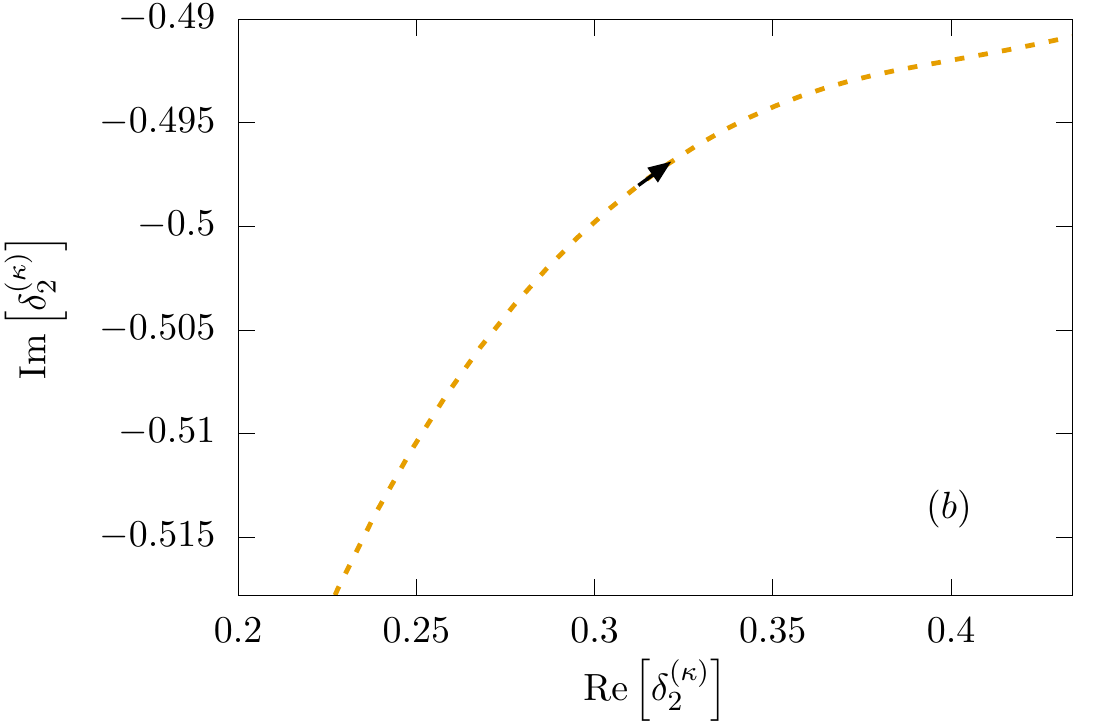}
	\end{tabular}
	\caption{Trajectories of the additional zeros of $\kappa(\lambda)$ in the physical strip, associated with the leading eigenvalue $\Lambda_0^t$ at small times. The figure corresponds to anisotropy $\Delta=1/2$, for which two auxiliary functions $y_1(\lambda)$ and $\kappa(\lambda)$ are introduced, satisfying the $Y$-system in Eqs.~\eqref{eq:y-system_p2_eq1} and \eqref{eq:y-system_p2_eq2}. Subfigure $(a)$: the plots correspond to the trajectory of the first additional zero of $\kappa(\lambda)$ from $t\simeq 1.25$ (at which it enters the physical strip) to $t\simeq 7.2$. Subfigure $(b)$: the plots corresponds to the trajectory of the second additional zero of $\kappa(\lambda)$ from $t\simeq 4.39$ (at which it enters the physical strip) to $t\simeq 7.2$. Arrows show the direction of the trajectories.}
	\label{fig:additional_zeros}
\end{figure}
The calculations for additional zeros of $y_1(\lambda)$ are exactly the same. One should only pay attention to the fact that now zeros of $y_1$ are of order $2$: if this was not the case, the function $1+\kappa(u)$ would display a point of non-analyticity. Again, up to a global additive constant, we get the additional term
\bea
-4\pi i\, {\rm sign}\left[\mathcal{I}_\delta^{(y)}\right] \left\{\mathcal{L}\left[\left(\delta^{(y)}-i{\rm sign}\left[\mathcal{I}_\delta^{(y)}\right]\gamma/2\right)-\lambda\right]\right.\nonumber \\
\left.+\mathcal{L}\left[\left(\delta^{(y)}-i{\rm sign}\left[\mathcal{I}_\delta^{(y)}\right]\gamma/2\right)+\lambda\right]\right\}\,.
\eea

We can collect these calculations and provide the final result for the integral equations in the presence of additional zeros. Suppose that $y_1(\lambda)$ and $\kappa(\lambda)$ have respectively $n^{y}$ and $n^{\kappa}$ additional zeros in the physical strip;  then, in the Trotter limit $N\to\infty$, we obtain the following set of TBA equations
\bea 
\ln y_1(\lambda) 
&=& - 2\pi i \sin(\gamma) t s(\lambda) -2\ln\left(\coth\frac{3\lambda}{2}\right) +2\sum_{j=1}^{n^{y}}\mathcal{G}\left(\lambda,\delta^{(y)}_j\right)\nonumber	\\
&+&2 s\ast \ln(1+\kappa )+\log C_1\,,\label{eq:TBA1}\\
\ln \kappa(\lambda) &=&  2\ln\left(\coth\frac{3\lambda}{2}\right)+ \sum_{j=1}^{n^{\kappa}}\mathcal{G}\left(\lambda,\delta^{(\kappa)}_j\right)+s\ast \ln(1+y_1 )+\log C_2\,,\label{eq:TBA2}
\eea  
where
\bea
\mathcal{G}(\lambda,\delta)&=&-2\pi i\, {\rm sign}\left[{\rm Im}\,\delta\right] \left\{\mathcal{L}\left[\left(\delta-i{\rm sign}\left[{\rm Im}\,\delta\right]\gamma/2\right)-\lambda\right]\right.\nonumber\\
&+&\left.\mathcal{L}\left[\left(\delta-i{\rm sign}\left[{\rm Im}\,\delta\right]\gamma/2\right)+\lambda\right]\right\}\,,
\label{eq:g_function}
\eea
while $C_1$ and $C_2$ are two constants which should be fixed for the particular eigenvalue investigated. Indeed, noticing that $\lim_{\lambda\to\infty}\mathcal{G}(\lambda,\delta)=0$, and defining
\bea
y_1(\infty)&=&\lim_{\lambda\to\infty}y_1(\lambda)\,,\\
\kappa(\infty)&=&\lim_{\lambda\to\infty}\kappa(\lambda)\,,
\eea
we obtain from Eq.~\eqref{eq:Y1limit}
\bea
C_1=\frac{y_1(\infty)}{1+\kappa(\infty)}\,,\label{eq:c1}\\
C_2=\frac{\kappa^2(\infty)}{1+y_1(\infty)}\,.\label{eq:c2}
\eea
In the same way, the equation for the eigenvalue of the transfer matrix has to be modified in the presence of additional zeros as
\be
\ln \Lambda(\lambda) = \sum_{j=1}^{n^{\kappa}}\mathcal{G}\left(\lambda,\delta^{(\kappa)}_j\right)+  s\ast \ln\left( \frac{1+y_1}{1+\tilde{y}_1} \right)    -  s \ast \psi_N \,.
\ee

Since the values of the zeros $\{\delta_j^{(y/\kappa)}\}$ are not known analytically, they need to be determined self-consistently. In particular, using the $Y$-system relations, they are immediately seen to satisfy
\bea
y_1\left(\delta^{(\kappa)}_j\pm i\frac{\gamma}{2}\right)&=&-1\,,\label{eq:zero_y}\\
\kappa\left(\delta^{(y)}_j\pm i\frac{\gamma}{2}\right)&=&-1\,.\label{eq:zero_k}
\eea
These equations complement those in \eqref{eq:TBA1} and \eqref{eq:TBA2}, and finally allow us to compute the real-time evolution of a given eigenvalue $\Lambda_\ell^{t}(\lambda)$. In order to obtain explicit numerical results, one can proceed as follows. First, one starts with an initial guess on the position of the additional zeros and poles $\{\delta_j^{(y/\kappa)}\}$. Using this guess, one solves the integral equations \eqref{eq:TBA1} and \eqref{eq:TBA2}, yielding an approximation for $y_1(\lambda)$ and $\kappa(\lambda)$. Next, employing the latter, one solves Eqs.~\eqref{eq:zero_y} and \eqref{eq:zero_k} for $\{\delta_j^{(y/\kappa)}\}$, which serve as an improved guess for the next iteration. 

\begin{figure}
	\begin{tabular}{lll}
		\includegraphics[scale=0.7]{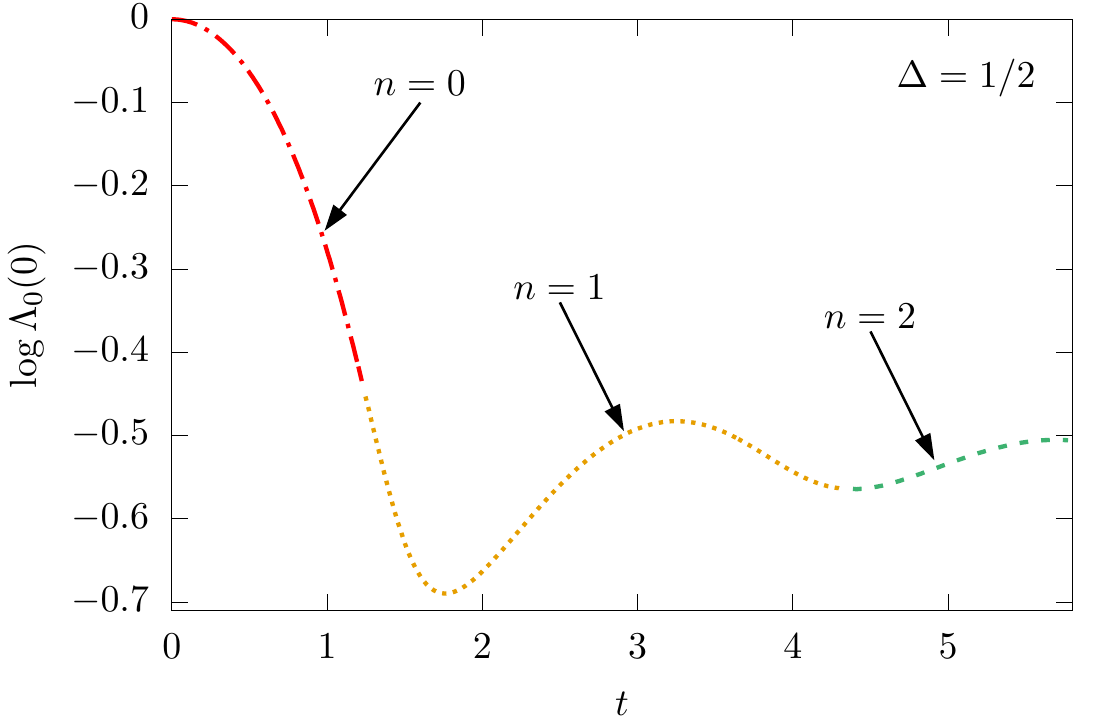} &
		\hspace{-0.12cm}\includegraphics[scale=0.7]{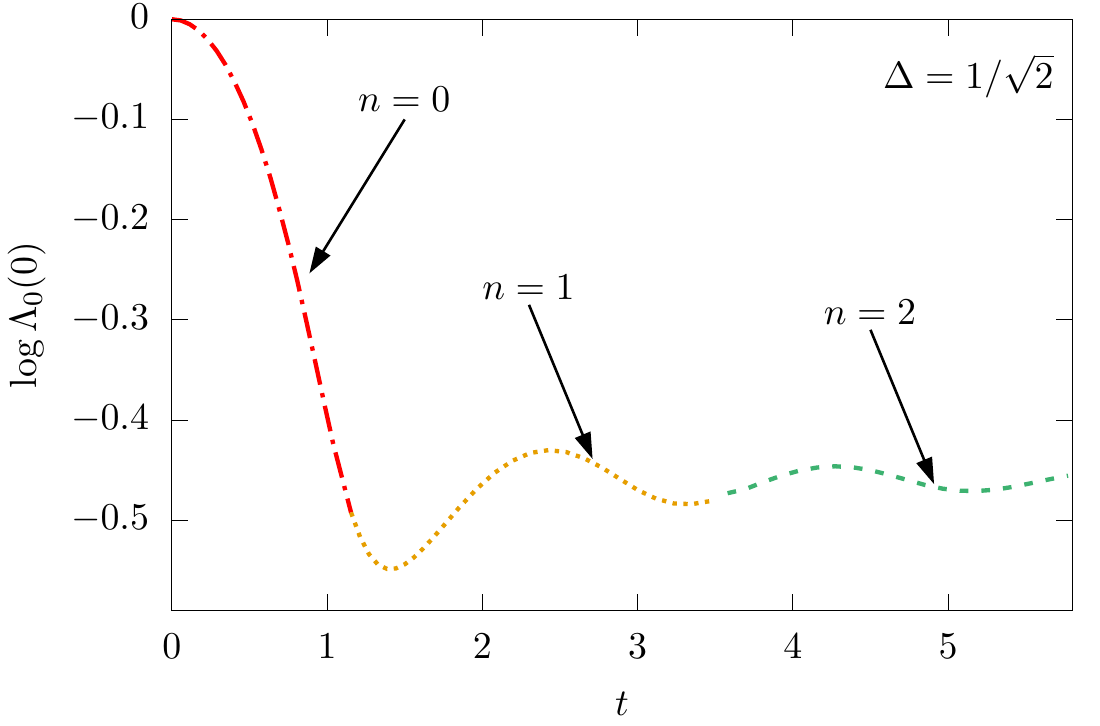}
	\end{tabular}
	\caption{Time-evolution of the logarithm of the eigenvalue $\Lambda^{t}_0(0)$, which is the leading one at short times. The two plots correspond to different values of the anisotropy. In the figures, we explicitly indicated the number of additional zeros of $\kappa(\lambda)$ entering the physical strip for each time interval (no additional zeros of $y_1(\lambda)$ are seen to appear).}
	\label{fig:lambda0}
\end{figure}

The additional zeros follow non-trivial trajectories in the physical strip, as displayed in Fig.~\ref{fig:additional_zeros}. Furthermore, their number can also vary in time. In Fig.~\ref{fig:lambda0} we display the leading eigenvalue $\Lambda^{t}_0$ for different values of the anisotropy as a function of time. In each plot, we also specify the number of additional singularities which enter the physical strip, and which have to be consistently determined from Eqs.~\eqref{eq:zero_y}, Eqs.~\eqref{eq:zero_k}.

From the numerical point of view, there is an additional non-trivial complication. Indeed, the driving term in Eq.~\eqref{eq:TBA1} is imaginary and one needs to be careful with the determination of the branch of the logarithm. In fact, in order to obtain a continuous solution to these equations, one can not avoid to consider the logarithm as a function defined on a multi-sheeted Riemann surface. Accordingly, $y_j(\lambda)$ and $\kappa(\lambda)$ need to be thought of as functions taking value in this surface. For completeness, a detailed discussion on this issue is reported in \ref{sec:riemann_sheet}, together with other technical aspects of the numerical solution to the non-linear integral equations.

\section{The full spectrum of the Quantum Transfer Matrix}
\label{sec:full_spectrum}

In the last sections, we have solved the problem of computing a single eigenvalue of the boundary transfer matrix for real times. In particular, we have followed the evolution of the eigenvalue $\Lambda^t_0$ which at $t=0$ is the leading one. As we have already mentioned, however, a crossing of eigenvalues will in general occur after a certain time $\bar{t}$: for $t>\bar{t}$ the eigenvalue $\Lambda^t_0$ will not be the leading one anymore.

As it should be clear from our discussion in the previous sections, the Bethe ansatz method allows us to follow the dynamic of a single eigenvalue continuously, starting from a given time $t$. Ideally, then, one should compute for a given time the full spectrum of the transfer matrix, so that one could keep track of each crossings of the eigenvalues at later times. One can summarize the procedure to do so, as follows:
\begin{itemize}
	\item diagonalize the transfer matrix at finite Trotter number;
	\item for each excitation, find the location of the additional zeros of the functions $y_j$ and $\kappa$;  
	\item use these as an input for the ``excited-state'' TBA procedure described in the previous section.
\end{itemize}

Let us follow these steps in detail for the first few leading states at time $t=0.7$. While the procedure works in principle for states with arbitrary values of the magnetization $S_z$, the leading QTM eigenvalue always appears to lie in the sector $S_z=0$ and we will therefore restrict to the latter in what follows. 
The time $t=0.7$ lies prior to any crossing, and the leading eigenstate is that studied in the previous section. The next leading states are characterized by a set of additional zeros in the physical strip, which we sum up in Table \ref{table:zeros}.
\begin{table}
	\centering
	\resizebox{\columnwidth}{!}{%
	\begin{tabular}{c | c | c| c }
		& $2N=6$ & $2N=8$ & $N \to \infty$
		\\
		\hline
		level 1 & &  &  \\
		\hline
		level 2 & $\kappa$  \hfill    & $\kappa$ &  $\kappa$ \\
		&  \small ${ \pm (0.252536+0.301549 i)}$  &   ${ \pm (0.252163+0.303160 i)}$  & $\pm( 0.25167 + 0.30521 i)$  \\
		&   \small  $\pm (0.009101-0.107342 i)$  & $\pm (0.008594-0.110679 i)$ & $\pm( 0.007978 - 0.11492 i)$ \\
		\hline
		level 3 & $\kappa$  \hfill    & $\kappa$ &  $\kappa$ \\
		&  \small ${ \pm (0.254856+0.233405 i)}$  &   ${ \pm (0.254385+0.232024 i)}$  & $\pm (0.253737+0.230304 i)$ \\
		&   \small  $\pm (0.013347+0.072526 i)$  & $\pm (0.012880+0.077114 i)$ &   $\pm (0.0123037+0.0828667 i)$\\
		& $y_1$  \hfill    & $y_1$ &  $y_1$ \\
		&  \small ${ \pm (0.034283-0.500868 i)[2]}$  &   ${ \pm (0.035795-0.499081 i)[2]}$  & ${ \pm (0.038054-0.496957 i)[2]}$  \\
		\hline
		level 4 & $\kappa$  \hfill    & $\kappa$ &  $\kappa$ \\
		&  \small ${ \pm (0.258274+0.273484 i)}$ &   ${ \pm (0.257894+0.276841i)}$  &  ${ \pm (0.257327+0.280913 i)}$ \\
		&   \small  $\pm (0.003950-0.034612 i)$  & $\pm (0.003278-0.042546 i)$ &  $\pm (0.0026001-0.052087 i)$ \\
		\hline
	\end{tabular}
	}
	\caption{Additional zeros [inside the physical strip \eqref{eq:physical_strip}] associated with the first leading eigenvalues of the boundary QTM in the zero magnetization sector, at time $t=0.7$ and for $p=2$. The multiplicity of zeros, when different from $1$, is indicated by brackets. The last column is obtained from the self-consistent solution to Eqs.~\eqref{eq:zero_y} and \eqref{eq:zero_k}.}
	\label{table:zeros}
\end{table}

\begin{figure}
	\centering
	\includegraphics[scale=0.95]{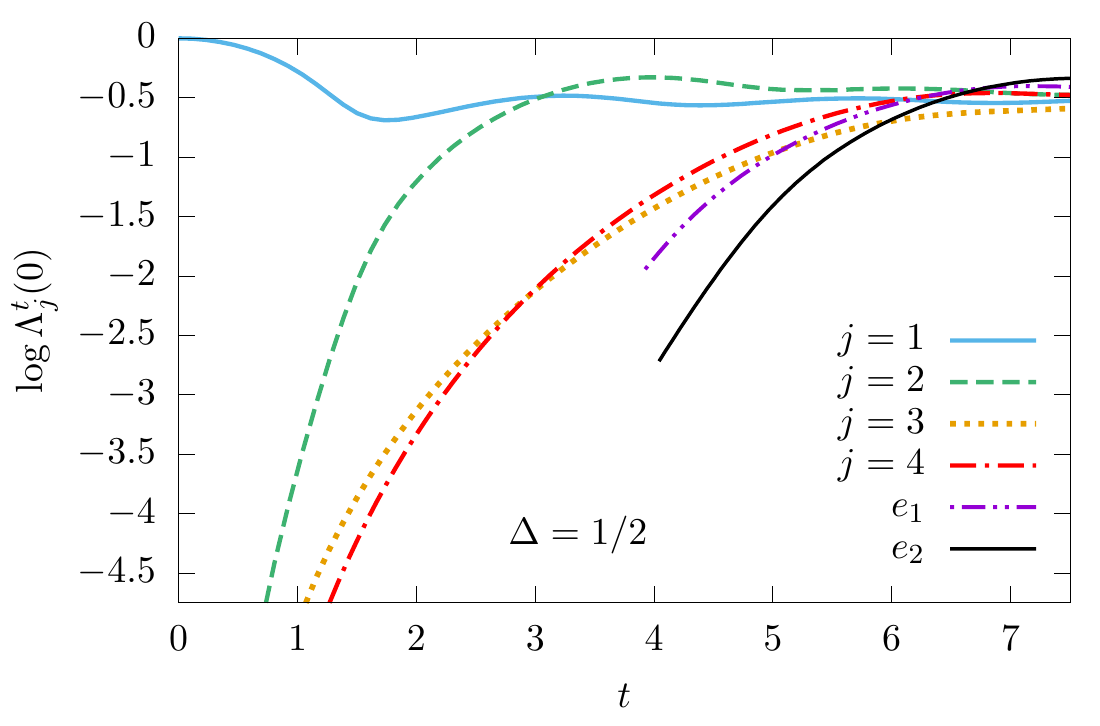}
	\caption{Time evolution of the spectrum of the boundary QTM in the sector of zero magnetization. The plot corresponds to anisotropy $\Delta=1/2$. The eigenvalues labeled with $j=1,\ldots, 4$ correspond to the first $4$ ones at time $t=0.7$. The eigenvalues labeled with $e_1$ and $e_2$ are involved in the crossings arising at later times. Their analytical structure is identified at $t=4$, for which they can be easily studied numerically as they do not lie too deep into the spectrum of the boundary QTM.}	
	\label{fig:final_losch1}
\end{figure}

The location of these additional zeros is quite stable upon increasing $N$, and can be reliably used as an input for the iterative scheme described in Sec.~\ref{sec:generaltimes}. The resulting eigenvalues are plotted in Fig.~\ref{fig:final_losch1}.
Importantly, these allow us to observe a first crossing between the levels $1$ and $2$ at time $t \simeq 3.05$, yielding a first point of non-analyticity of the Loschmidt echo. 

In order to observe crossings at later times, one could in principle follow the same approach for further excited states. However, it turns out that this is not convenient, as the states involved in crossings at later times are found to lie rather deep in the spectrum of the relevant boundary transfer matrix for $t=0.7$, and are therefore difficult to identify systematically. Accordingly, we proceed in a more pragmatic way. In particular, we study the boundary QTM spectrum as a function of $t$ for finite sizes $2N=6,8,10$, and identify the states such that the associated eigenvalues become the leading one within a finite given time window. In this way we managed to identify the states involved in two subsequent crossings, for which we characterized the additional zeros at a time $t=4$. Next, we solve the resulting integral equations to arbitrary times. The final result of this procedure is shown in Fig.~\ref{fig:final_losch1}. 

\begin{figure}
	\begin{tabular}{lll}
		\includegraphics[scale=0.7]{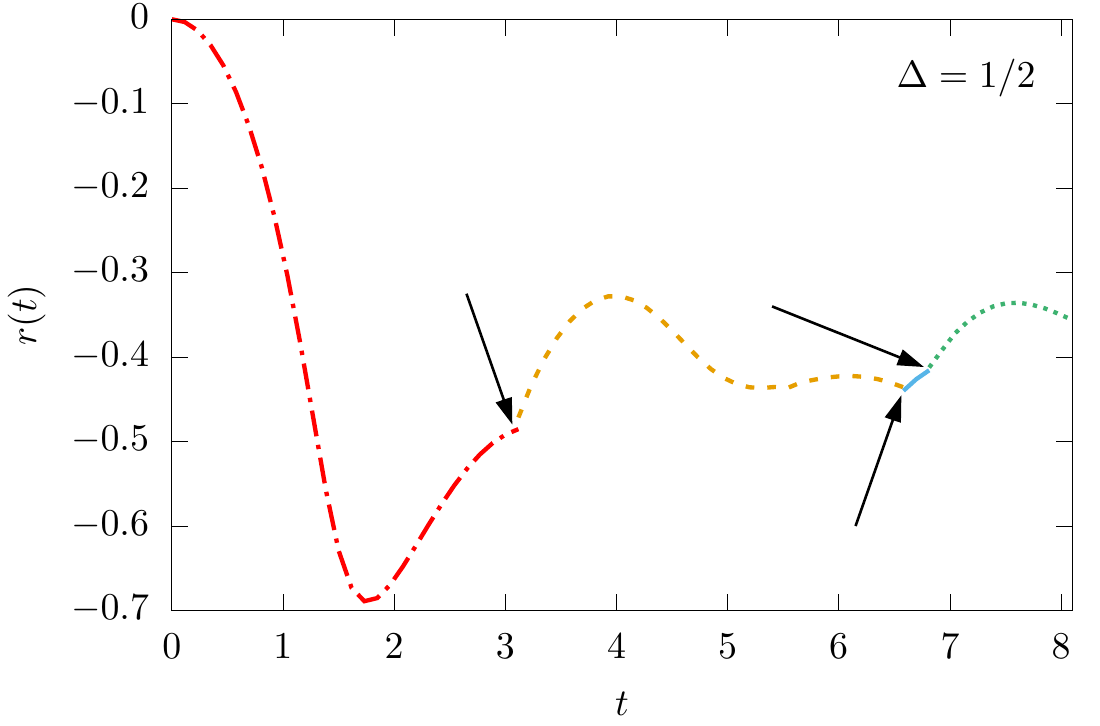} &  \hspace{-0.2cm}\includegraphics[scale=0.7]{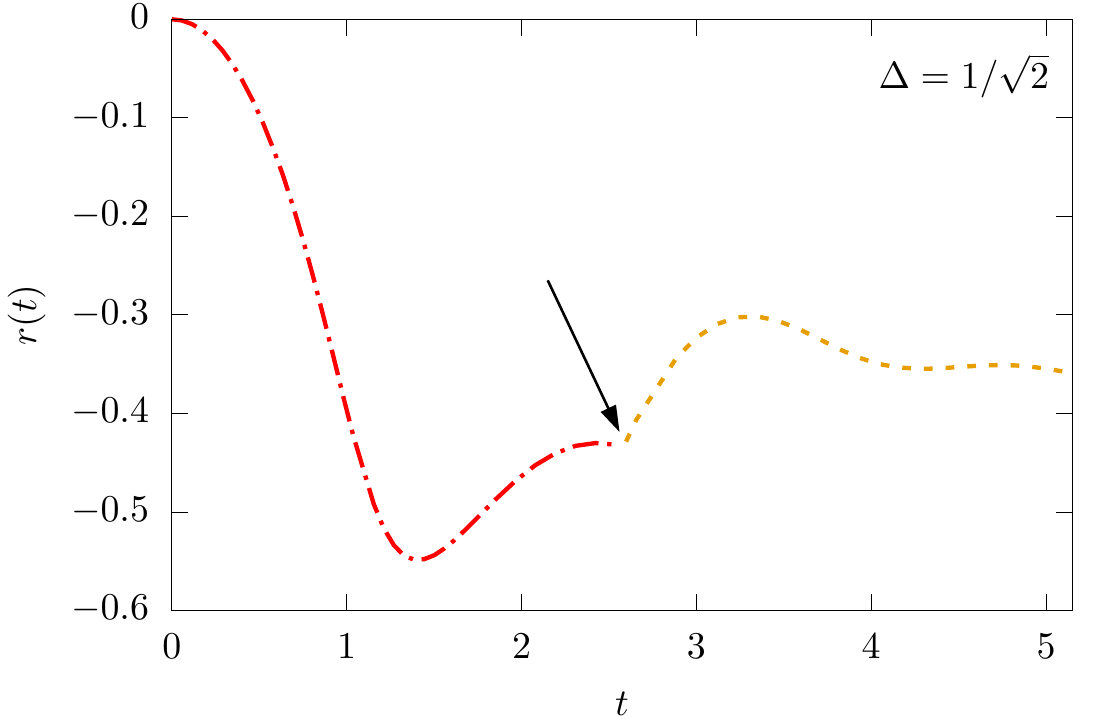}
	\end{tabular}
	\caption{Exact return rate as a function of time, as computed using the techniques detailed in Sec.~\ref{sec:generaltimes} and \ref{sec:full_spectrum}. The plots correspond to anisotropies $\Delta=1/2$ and $\Delta=1/\sqrt{2}$. The points of non-analyticities are explicitly highlighted with arrows, and correspond to crossings of the eigenvalues of the boundary transfer matrix, \emph{cf.} Fig.~\ref{fig:final_losch1}. Different colors correspond to the fact that the return rate is determined by different eigenvalues which become the leading ones at different times.
	}
	\label{fig:final_losch2}
\end{figure}

By selecting at each time the leading eigenvalue, one is left with the final exact result for the return rate, and hence the Loschmidt echo per site. This is shown in Fig.~\ref{fig:final_losch2}, for different values of the anisotropy. Our results were tested against iTEBD simulations \cite{Vida07}, and calculations from exact diagonalization at finite system size, displaying perfect agreement. As time is increased further, several additional points of non-analyticities are expected to arise; these should correspond to eigenvalues lying deeper and deeper in the spectrum at smaller time. This is in fact a limitation of our method, as these states become increasingly difficult to track in the spectrum of the QTM at finite $N$.
In order to tackle arbitrary time and, in particular, the problem of the asymptotics, it would be much more satisfactory to have at hand a set of integral equations incorporating in a self-consistent way the analytical properties of the leading eigenvalue throughout its crossings. While we have not been able to achieve this goal at present, we hope to return to these issues in future works.

Before closing this section, we point out a rather remarkable feature that we have observed, namely that all of the crossings involving the leading eigenvalue seem to coincide exactly with a change in the analytic structure of the $y$ and $\kappa$ functions. For instance, precisely at the location of the first crossing between the levels $1$ and $2$, the number of additional pairs of zeros of $\kappa$ for the level $1$ changes from zero to $1$. We were not able to provide a theoretical justification for this phenomenon, and at this stage we report it as a simple observation.

\section{Conclusion}
\label{sec:conclusion}

We addressed the computation of the Loschmidt echo in the XXZ spin-$1/2$ chain, for a special class of integrable initial states \cite{PiPV17_int}. By employing a QTM approach, we have provided an analytic solution at real times, completing the programme initiated in \cite{Pozs13,PiPV17}. Within our method, the Loschmidt echo is obtained from the leading eigenvalue of an appropriate boundary QTM. As the time increases, crossings occur giving rise to points of non-analyticity, which are fully captured by our techniques.

Although our approach could be in principle used to study the full spectrum of the boundary QTM, one is practically limited in the number of eigenvalues which can be computed. This results in a limitation in the time interval which can be considered: indeed, as time increases, more and more crossings are expected, and a very large number of eigenvalues should be taken into account. In particular, using our method, we do not have access to the regime of infinite times, and the study of the asymptotic behavior of the Loschmidt echo remains an interesting open question to be investigated.

Our calculations show that TBA techniques, traditionally tailored for
thermal physics, can be successfully used to obtain explicit analytic
results also for real-time dynamics. Hence, our approach constitutes a
promising direction towards the ambitious goal of computing the time
evolution of local observables after a quantum quench. Applications of
the techniques employed in this work to this very important question
are currently under investigation.

\section{Acknowledgments}

We are very grateful to Mario Collura for providing us with iTEBD data, to test our numerical scheme for evaluation of the analytic formulas derived in this work. We acknowledge useful discussions with Pasquale Calabrese, Patrick Dorey, Fabian Essler, G\'abor T\'akacs and Roberto Tateo. B.P. acknowledges support from the ``Premium'' Postdoctoral
Program of the Hungarian Academy of Sciences, the K2016 grant no. 119204 and the KH-17 grant no. 125567 of the research agency NKFIH. E.V. acknowledges support by the EPSRC under grant EP/N01930X.

\appendix

\section{The Riemann sheet TBA}
\label{sec:riemann_sheet}

We wish to illustrate the numerical solution to the TBA equation \eqref{eq:TBA1} and \eqref{eq:TBA2} for $t\in \mathbb{R}$. In order to do this, we consider the explicit case of the leading eigenvalue of the transfer matrix. For $t$ sufficiently small, there are no additional zeros; the TBA equations then read 
\bea 
\ln y_1 
&=& - 2\pi i t \sin(\gamma) s(\lambda) -2\ln\left(\coth\frac{3\lambda}{2}\right) +2 s\ast \ln(1+\kappa )\,,\label{eq:explicit_TBA1}\\
\ln \kappa &=&  2\ln\left(\coth\frac{3\lambda}{2}\right)+s\ast \ln(1+y_1 )\,.\label{eq:explicit_TBA2}
\eea 
It is convenient to employ the parametrization
\bea
y_1(\lambda)&=&\rho_{y}(\lambda)e^{i\varphi_y(\lambda)}\,,\\
\kappa(\lambda)&=&\rho_{\kappa}(\lambda)e^{i\varphi_\kappa(\lambda)}\,.
\eea
Furthermore, for reasons that will be clear later, it is convenient to introduce also the following parametrization
\bea
1+\kappa(\lambda)=R_{\kappa}(\lambda)e^{i\Phi_\kappa(\lambda)}\,,\label{eq:par_oneplusK}\\
1+y_1(\lambda)=R_{y}(\lambda)e^{i\Phi_y(\lambda)}\,.\label{eq:par_oneplusY}
\eea
All the functions $\rho_y(\lambda)$, $R_y(\lambda)$, $\varphi_y(\lambda)$, $\Phi_y(\lambda)$, are real for real values of $\lambda$. The integral equations above can be rewritten as
\bea
\ln \rho_y &=& 2 s\ast \ln R_\kappa 
-  \ln \left( \coth^2 \frac{3  \lambda}{2} \right)\,,\label{eq:NLIE1}\\
\varphi_y &=& - 2\pi \sin(\gamma) t\, s(\lambda)+2 s\ast \Phi_\kappa
\,,\label{eq:NLIE2}\\
\ln \rho_\kappa &=& s\ast \ln R_y
+  \ln \left( \coth^2 \frac{3 \lambda}{2} \right)\,,\label{eq:NLIE3}\\
\varphi_\kappa &=&  s\ast \Phi_y  \,.
\label{eq:NLIE4}
\eea
The system above consists of $4$ equations with $8$ unknown functions, so it can not be solved unless additional constraints are imposed. Indeed, the functions $R_{\kappa/y}(\lambda)$ and $\Phi_{\kappa/y}(\lambda)$ are not independent from $\rho_{\kappa/y}(\lambda)$ and $\phi_{\kappa/y}(\lambda)$. However, we argue that the dependence is non-trivial and in general ``non-local'': in order to obtain $\Phi_{\kappa/y}$ for a given $\lambda$ it is not enough to know the value taken by functions $\rho_{\kappa/y}$ and $\phi_{\kappa/y}$ at the same $\lambda$, but additional, non-local information should be provided.

To understand this better, we consider a very simple example. Suppose we assign the functions
\be
	\varphi(\lambda)=2\pi \lambda\,,\qquad \rho_y(\lambda)= \lambda\,.
	\label{eq:curve1}
\ee
As $\lambda$ varies, one can follow the evolution of $1+y_1(\lambda)$, and compute $R_{y}(\lambda)$ and $\Phi_y(\lambda)$ [defined in \eqref{eq:par_oneplusY}] accordingly. Note that if we do this naively, for example choosing
\be
\Phi_y(\lambda)=\log\left(1+y_1(\lambda)\right)\,,
\ee
with a fixed branch cut for the logarithm (for example $(-\infty,0]$), we obtain a discontinuous function $\Phi_y(\lambda)$: a discontinuity arises every time $1+y_{1}(\lambda)$ crosses the branch cut $(-\infty,0]$. On the other hand, we assume that the correct solutions to \eqref{eq:NLIE1}-\eqref{eq:NLIE4} are regular functions, so that no discontinuity should arise. 

In order to obtain continuous solutions $R_y(\lambda)$, $\Phi_y(\lambda)$, we introduce an infinitely sheeted Riemann surface, with a branch cut on the semi-infinite line $(-\infty,0]$. Then, we follow the curve $1+y_1(\lambda)$ on such a Riemann surface, and compute $R_y(\lambda)$ and $\Phi_y(\lambda)$ accordingly.
\begin{figure}
	\begin{tabular}{lll}
		\hspace{-0.7cm}\includegraphics[width=0.36\textwidth,height=0.27\textwidth]{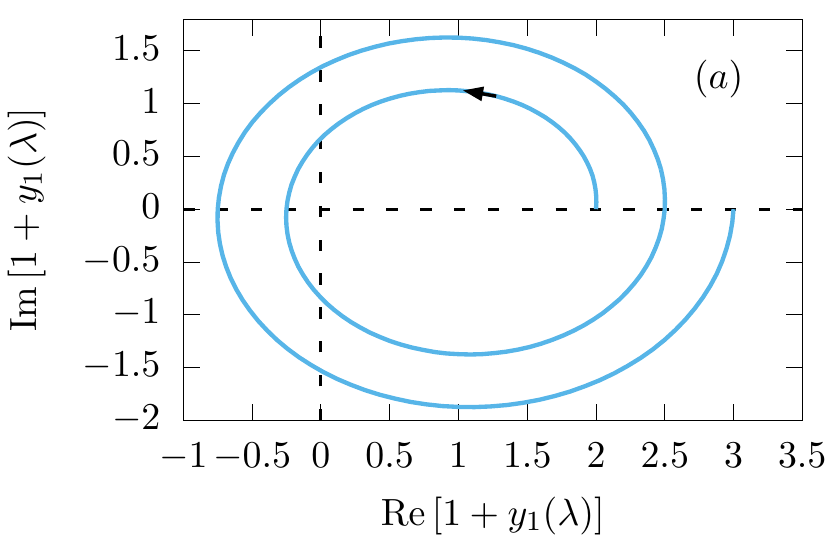} & 
		\hspace{-0.8cm}\includegraphics[width=0.36\textwidth,height=0.27\textwidth]{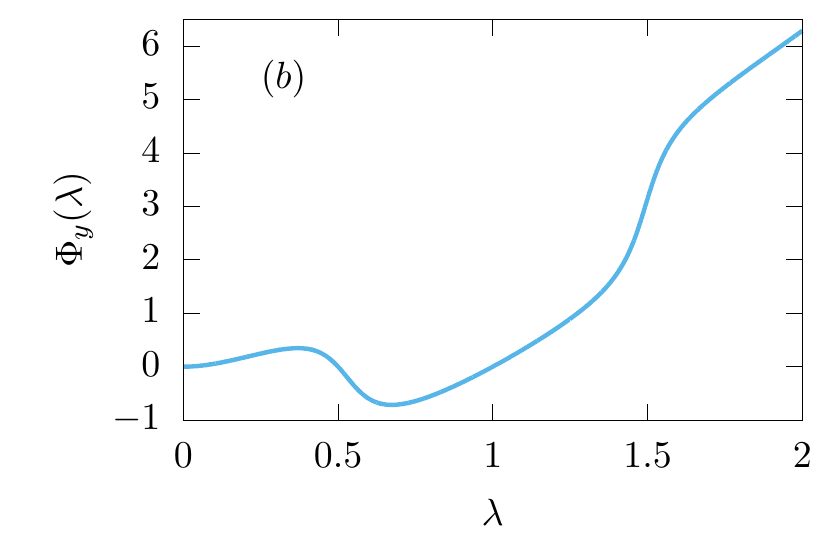} &  \hspace{-0.7cm}\includegraphics[width=0.36\textwidth,height=0.27\textwidth]{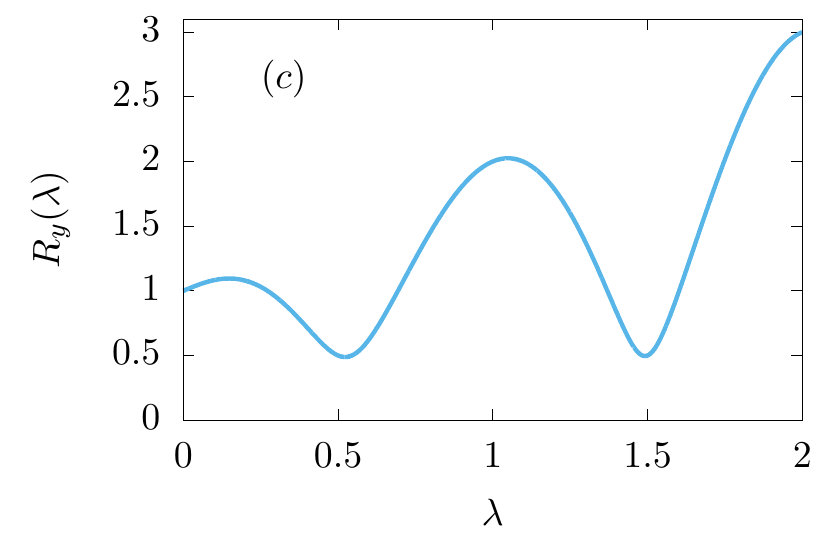}
	\end{tabular}
	\caption{Evolution (as a function of $\lambda$) of different curves corresponding to \eqref{eq:curve1}. In subfigure $(a)$ we plot the projection of the curve $1+y_1(\lambda)$ on a single sheet of the Riemann surface, as $\lambda$ is increased from $\lambda=0$ to $\lambda= 2$. In subfigures $(b)$ and $(c)$ we report the functions $\Phi_y(\lambda)$, $R_y(\lambda)$ which are computed by following the evolution of $1+y_1(\lambda)$ as $\lambda$ increases. }
	\label{fig:Curve1}
\end{figure}
This strategy is displayed in Fig.~\ref{fig:Curve1}. In subfigures $(a)$ we plot the projection of the curve $1+y_1(\lambda)$ on a single sheet of the Riemann surface, as $\lambda$ is increased from $\lambda=0$ to $\lambda= 2$. By following the curve, one can compute continuous functions $\Phi_y(\lambda)$ and $R_{y}(\lambda)$ which satisfy \eqref{eq:par_oneplusY}. In subfigures $(b)$ and $(c)$ we report the functions $\Phi_y(\lambda)$, $R_y(\lambda)$ computed in this was.

\begin{figure}
	\begin{tabular}{lll}
		\hspace{-0.7cm}\includegraphics[width=0.36\textwidth,height=0.27\textwidth]{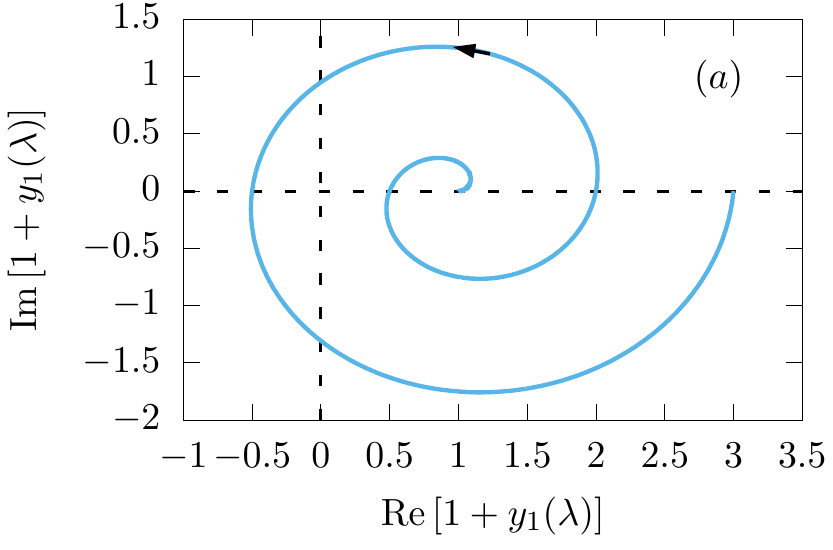} & 
		\hspace{-0.8cm}\includegraphics[width=0.36\textwidth,height=0.27\textwidth]{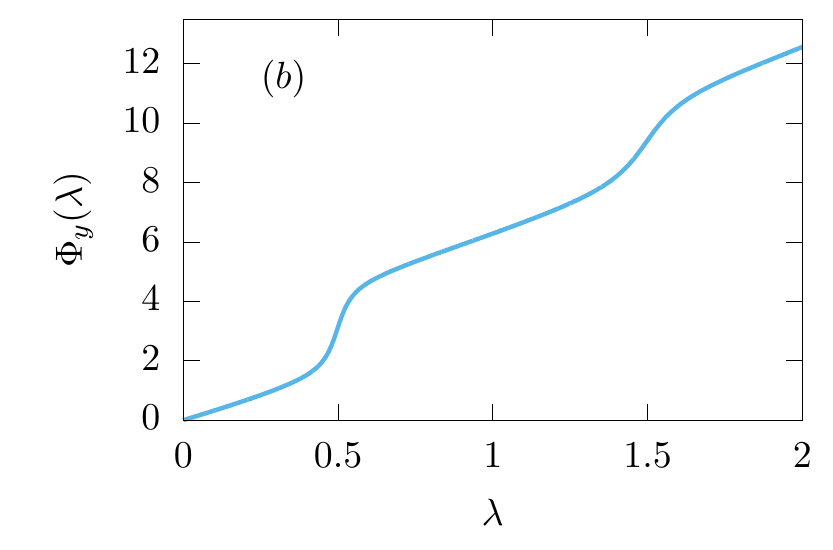} &  \hspace{-0.7cm}\includegraphics[width=0.36\textwidth,height=0.27\textwidth]{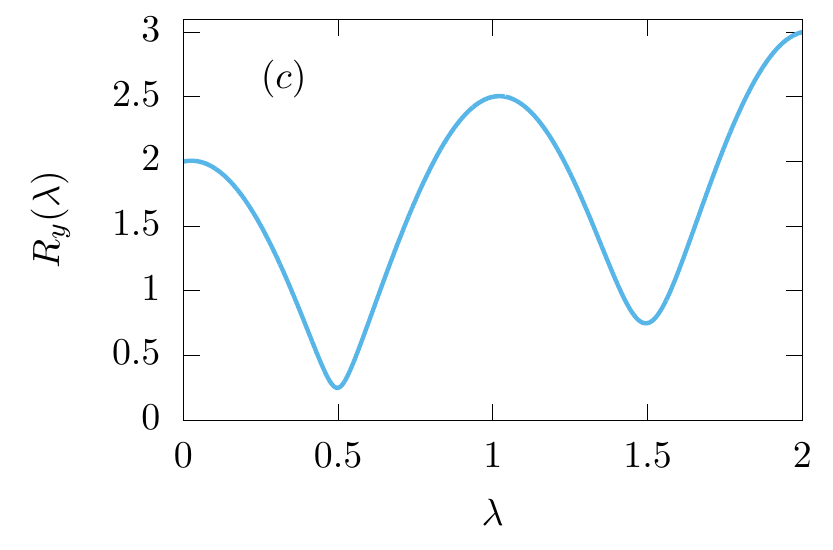}
	\end{tabular}
	\caption{Evolution (as a function of $\lambda$) of different curves corresponding to \eqref{eq:curve2}. In subfigure $(a)$ we plot the projection of the curve $1+y_1(\lambda)$ on a single sheet of the Riemann surface, as $\lambda$ is increased from $\lambda=0$ to $\lambda= 2$. In subfigures $(b)$ and $(c)$ we report the functions $\Phi_y(\lambda)$, $R_y(\lambda)$ which are computed by following the evolution of $1+y_1(\lambda)$ as $\lambda$ increases.  }
	\label{fig:Curve2}
\end{figure}

To see that the value of $\Phi_y(\lambda)$ does not uniquely depend on the value of $\rho_y$ and $\varphi_y$ computed in $\lambda$, consider a different curve
\be
\tilde{\varphi}_y(\lambda)=2\pi \lambda\,,\qquad \tilde{\rho}_y(\lambda)= 1+\frac{1}{2}\lambda\,.
\label{eq:curve2}
\ee
Once again, as $\lambda$ varies we obtain a curve in the infinitely sheeted Riemann surface corresponding to $1+y_1(\lambda)$. Its projection on a single Riemann sheet is reported in subfigure $(a)$ of Fig.~\ref{fig:Curve2}, while the corresponding continuous functions $\Phi_y(\lambda)$ and $\rho_y(\lambda)$ are displayed in subfigures $(b)$ and $(c)$. Importantly, we see that 
\bea
\varphi_y(\lambda=2)&=&\tilde{\varphi}_y(\lambda=2)=4\pi\,,\\
\rho_y(\lambda=2)&=&\tilde{\rho}_y(\lambda=2)=2\,.
\eea
However, we have
\bea
\Phi_y(\lambda=2)&=&2\pi\,,\\
\tilde{\Phi}_y(\lambda=2)&=&4\pi\,.
\eea

From this discussion it follows that, in order to find regular solutions of the system \eqref{eq:NLIE1}-\eqref{eq:NLIE4}, one needs to take explicitly into account an infinitely-sheeted Riemann surface. In practice, we use the following numerical scheme:
\begin{enumerate}
	\item Initialize the values of $R_{\kappa/y}$ and $\Phi_{\kappa,y}$.
	\item Compute directly the value of $\rho_y$ and $\varphi_y$ from \eqref{eq:NLIE1}--\eqref{eq:NLIE4}. \label{item:step2}
	\item Given the functions $\rho_{\kappa/y}(\lambda)$ and $\varphi_{\kappa,y}(\lambda)$, compute $R_{y/\kappa}(\lambda)$ and $\Phi_{y/\kappa}$ following the curve $1+y_1(\lambda)$ on the multi-sheeted Riemann surface. \label{item:step3}
	\item Repeat steps \eqref{item:step2} and \eqref{item:step3} until convergence is reached.
\end{enumerate}
The most delicate step in this scheme is \eqref{item:step3}, but can nevertheless be implemented numerically. We now discuss results for the solution to \eqref{eq:NLIE1}-\eqref{eq:NLIE4}.

\begin{figure}
	\centering
	\begin{tabular}{cccc}
		\includegraphics[scale=0.7]{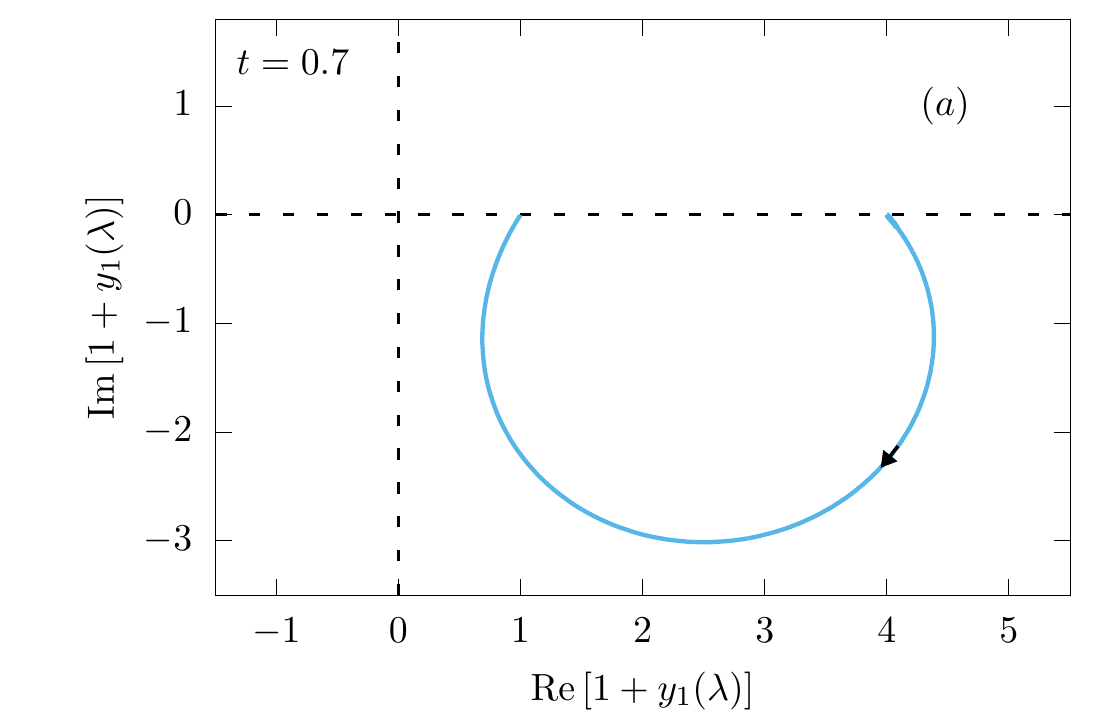} & 
		\hspace{-0.7cm}\includegraphics[scale=0.7]{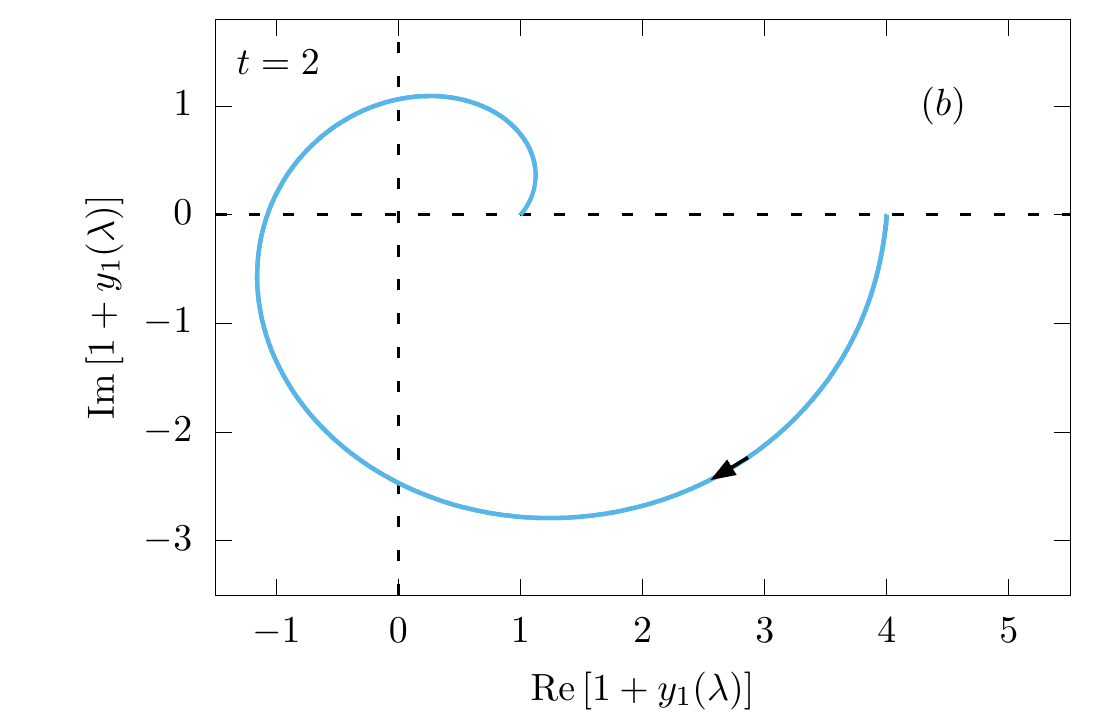} & \\ \includegraphics[scale=0.7]{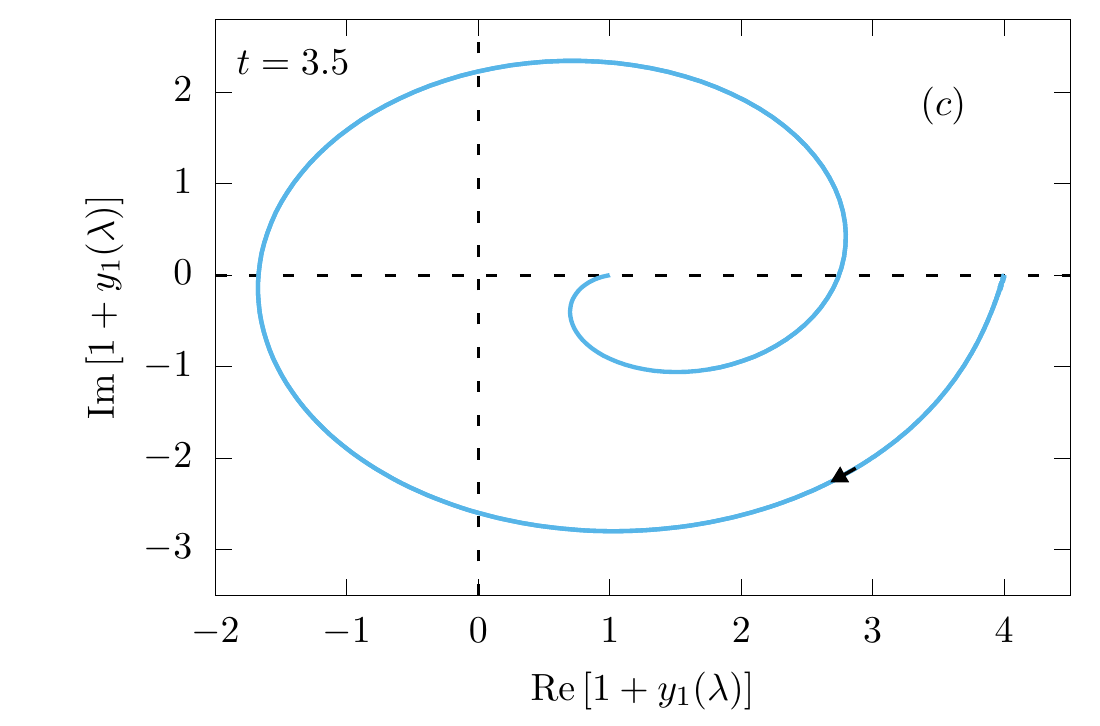}&
		\hspace{-0.7cm}\includegraphics[scale=0.7]{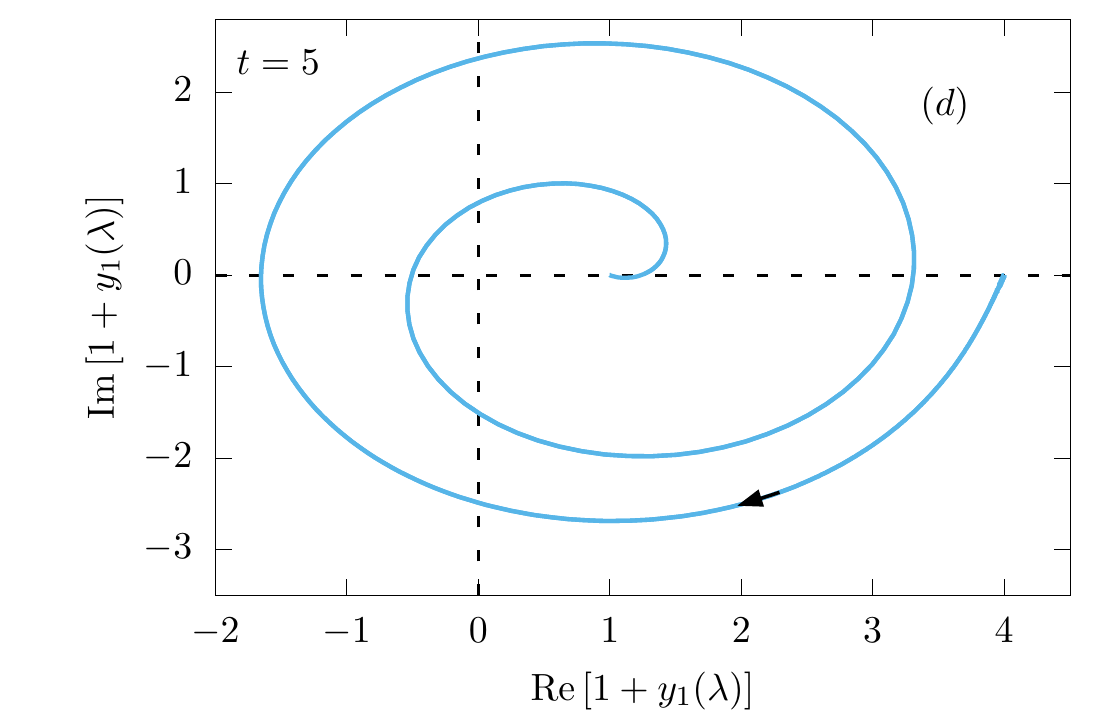}&
	\end{tabular}
	\caption{Evolution (as a function of $\lambda$) of the actual solution to the Eqs.~\eqref{eq:NLIE1}-\eqref{eq:NLIE4}. In each subfigure we plot the projection of $1+y_1(\lambda)$ on a single sheet, as $\lambda$ is increased from $\lambda=-10$ to $\lambda= 0$. The plots correspond to increasing times: $t=0.7, 2, 3.5,5$.}
	\label{fig:actual_Curves}
\end{figure}

\begin{figure}
	\centering
	\includegraphics[scale=0.78]{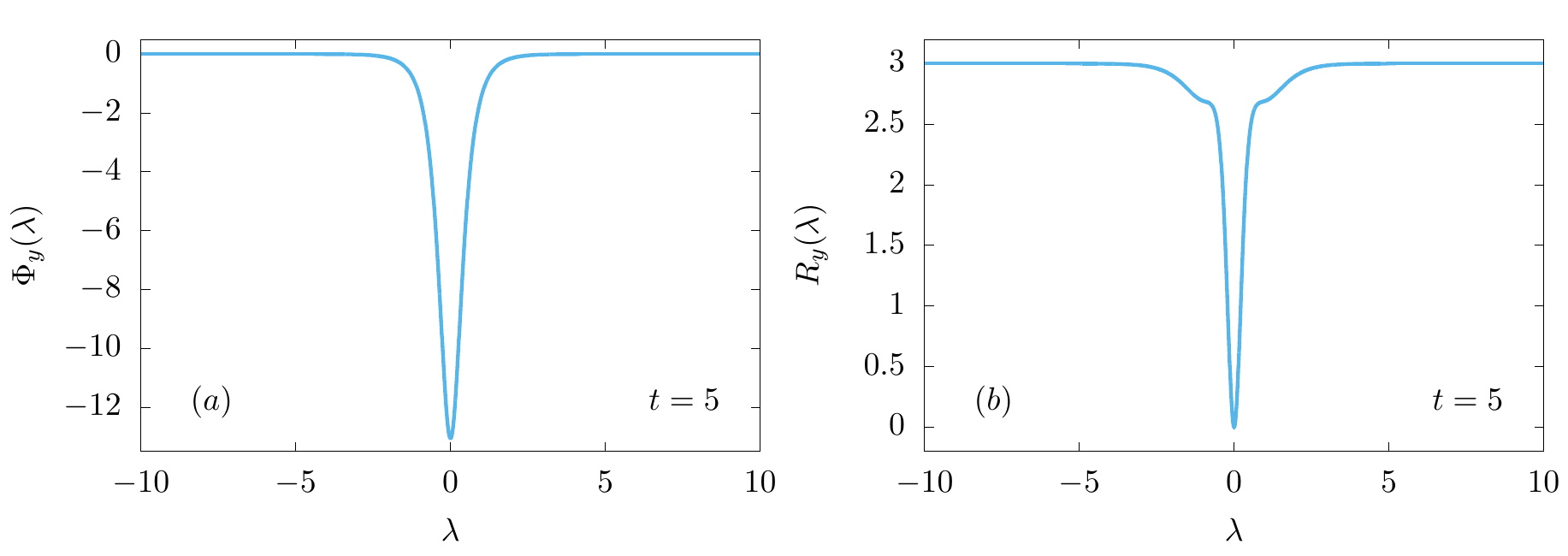} 
	\caption{Solutions $\Phi_y(\lambda)$ and $R_y(\lambda)$ to the Eqs.~\eqref{eq:NLIE1}-\eqref{eq:NLIE4} for $t=5$.}
	\label{fig:actual_abs_arg}
\end{figure}

We report in Fig.~\ref{fig:actual_Curves} numerical data for the solution to Eqs.~\eqref{eq:NLIE1}-\eqref{eq:NLIE4} corresponding to four different times. In particular, subfigures $(a)$, $(b)$, $(c)$, and $(d)$ correspond to times $t=0.7,2, 3.5, 5$. In each subfigure, we report the curve $1+y_1(\lambda)$ obtained by increasing the parameter $\lambda$ from $\lambda=-10$ to $\lambda=0$.  We see that for small times ($t=0.7$) the curve $1+y_1(\lambda)$ lies entirely in one single Riemann sheet, and a solution to Eqs.~\eqref{eq:NLIE1}-\eqref{eq:NLIE4} is straightforward. As time increases, the same curve eventually enters a new Riemann sheet. For example, for $t=2$ [subfigure $(b)$] the curve crosses the branch cut $(-\infty,0]$ once, and enters into a new Riemann sheet. We see that, as the curve $1+y_1(\lambda)$ in general wraps around the origin several times, it is not possible to choose a branch cut of the logarithm such that the latter lives on a single Riemann sheet, and necessarily a multi-sheeted surface has to be introduced. This could also be seen from Fig.~\ref{fig:actual_abs_arg}, where we plot the functions $\Phi_y(\lambda)$ and $R_y(\lambda)$ at time $t=5$.

It is important to note that at each time such that the curve $1+y_1(\lambda)$ enters a new Riemann sheet, a new zero of $\kappa$ appears in the physical strip, and hence an additional driving term has to be included in the equations. For example, in the particular case of the plot reported in subfigure $(b)$ of Fig.~\ref{fig:actual_Curves} the integral equations become
\bea
\ln \rho_y &=& 2 s\ast \ln R_\kappa 
-  \ln \left( \coth^2 \frac{3  \lambda}{2} \right)+\mathfrak{Re}\left[\mathcal{G}\left(\lambda,\delta^{(y)}\right)\right]\,,\label{eq:newNLIE1}\\
\varphi_y &=& - 2\pi \sin(\gamma) t\, s(\lambda)+2 s\ast \Phi_\kappa+\mathfrak{Im}\left[\mathcal{G}\left(\lambda,\delta^{(y)}\right)\right]
\,,\label{eq:newNLIE2}\\
\ln \rho_\kappa &=& s\ast \ln R_y
+  \ln \left( \coth^2 \frac{3 \lambda}{2} \right)\,,\label{eq:newNLIE3}\\
\varphi_\kappa &=&  s\ast \Phi_y  \,,
\label{eq:newNLIE4}
\eea
where $\mathcal{G}(\lambda,\delta^{(k)})$ is given in \eqref{eq:g_function}, while $\delta^{(k)}$ has to be computed self-consistently by solving Eq.~\eqref{eq:zero_k}.

Note that the numerical iteration to compute the position of the additional zeros is rather stable and efficient, especially when an initial reasonable guess is given. In fact, it only becomes delicate in correspondence of those times for which a new zero enters the physical strip and a new contribution has to be taken into account; in these cases, care must be taken to follow continuously the solution as time is increased slowly, and to provide an accurate initial guess for the position of the additional zeros.


\Bibliography{100}

\addcontentsline{toc}{section}{References}

\bibitem{CaEM16} 
P. Calabrese, F. H. L. Essler, and G. Mussardo, 
\href{http://dx.doi.org/10.1088/1742-5468/2016/06/064001}{J. Stat. Mech. (2016) 64001}.

\bibitem{cc-06}
P. Calabrese and J. Cardy, 
\href{http://dx.doi.org/10.1103/PhysRevLett.96.136801}{Phys. Rev. Lett. {\bf 96}, 136801 (2006)};\\
P. Calabrese and J. Cardy, 
\href{http://dx.doi.org/10.1088/1742-5468/2007/06/P06008}{J. Stat. Mech. (2007) P06008}.



\bibitem{PSSV11} 
A. Polkovnikov, K. Sengupta, A. Silva, and M. Vengalattore, 
\href{http://dx.doi.org/10.1103/RevModPhys.83.863}{Rev. Mod. Phys. {\bf 83}, 863 (2011)}.

\bibitem{ViRi16} 
L. Vidmar and M. Rigol, 
\href{http://dx.doi.org/10.1088/1742-5468/2016/06/064007}{J. Stat. Mech. (2016) 64007}.

\bibitem{EsFa16} 
F. H. L. Essler and M. Fagotti, 
\href{http://dx.doi.org/10.1088/1742-5468/2016/06/064002}{J. Stat. Mech. (2016) 64002}.

\bibitem{BeDo16Review} 
D. Bernard and B. Doyon, 
\href{http://dx.doi.org/10.1088/1742-5468/2016/06/064005}{J. Stat. Mech. (2016) 64005}.

\bibitem{VMreview}
R. Vasseur and J. E. Moore, \href{https://doi.org/10.1088/1742-5468/2016/06/064010}{J. Stat. Mech. (2016) 064010}.


\bibitem{RiDO08} 
M. Rigol, V. Dunjko, and M. Olshanii, 
\href{http://dx.doi.org/10.1038/nature06838}{Nature {\bf 452}, 854 (2008)}.

\bibitem{RDYO07} 
M. Rigol, V. Dunjko, V. Yurovsky, and M. Olshanii, 
\href{http://dx.doi.org/10.1103/PhysRevLett.98.050405}{Phys. Rev. Lett. {\bf 98}, 50405 (2007)}.

\bibitem{CaEF11} 
P. Calabrese, F. H. L. Essler, and M. Fagotti, 
\href{http://dx.doi.org/10.1103/PhysRevLett.106.227203}{Phys. Rev. Lett. {\bf 106}, 227203 (2011)};\\
P. Calabrese, F. H. L. Essler, and M. Fagotti, 
\href{http://dx.doi.org/10.1088/1742-5468/2012/07/P07022}{J. Stat. Mech. (2012) P07022};\\
P. Calabrese, F. H. L. Essler, and M. Fagotti, 
\href{http://dx.doi.org/10.1088/1742-5468/2012/07/P07016}{J. Stat. Mech. (2012) P07016}.

\bibitem{FaEs13} 
M. Fagotti and F. H. L. Essler, 
\href{http://dx.doi.org/10.1088/1742-5468/2013/07/P07012}{J. Stat. Mech. (2013) P07012};\\
M. Fagotti, M. Collura, F. H. L. Essler, and P. Calabrese, 
\href{http://dx.doi.org/10.1103/PhysRevB.89.125101}{Phys. Rev. B {\bf 89}, 125101 (2014)}.

\bibitem{Pozs13GGE} 
B. Pozsgay, 
\href{http://dx.doi.org/10.1088/1742-5468/2013/07/P07003}{J. Stat. Mech. (2013) P07003}.

\bibitem{IDWC15} 
E. Ilievski, J. De Nardis, B. Wouters, J.-S. Caux, F. H. L. Essler, and T. Prosen, 
\href{http://dx.doi.org/10.1103/PhysRevLett.115.157201}{Phys. Rev. Lett. {\bf 115}, 157201 (2015)}.

\bibitem{PiVC16} 
L. Piroli, E. Vernier, and P. Calabrese, 
\href{http://dx.doi.org/10.1103/PhysRevB.94.054313}{Phys. Rev. B {\bf 94}, 54313 (2016)};\\
L. Piroli, E. Vernier, P. Calabrese, and M. Rigol, 
\href{http://dx.doi.org/10.1103/PhysRevB.95.054308}{Phys. Rev. B {\bf 95}, 54308 (2017)}.

\bibitem{IlQC17} 
E. Ilievski, E. Quinn, and J.-S. Caux, 
\href{http://dx.doi.org/10.1103/PhysRevB.95.115128}{Phys. Rev. B {\bf 95}, 115128 (2017)}.

\bibitem{PoVW17} 
B. Pozsgay, E. Vernier, and M. A. Werner, 
\href{http://dx.doi.org/10.1088/1742-5468/aa82c1}{J. Stat. Mech. (2017) 93103}.

\bibitem{IMPZ16} 
E. Ilievski, M. Medenjak, T. Prosen, and L. Zadnik, 
\href{http://dx.doi.org/10.1088/1742-5468/2016/06/064008}{J. Stat. Mech. (2016) 64008}.



\bibitem{CaEs13} 
J.-S. Caux and F. H. L. Essler, 
\href{http://dx.doi.org/10.1103/PhysRevLett.110.257203}{Phys. Rev. Lett. {\bf 110}, 257203 (2013)}.

\bibitem{Caux16} 
J.-S. Caux, 
\href{http://dx.doi.org/10.1088/1742-5468/2016/06/064006}{J. Stat. Mech. (2016) 64006}.


\bibitem{BeSE14} 
B. Bertini, D. Schuricht, and F. H. L. Essler, 
\href{http://dx.doi.org/10.1088/1742-5468/2014/10/P10035}{J. Stat. Mech. (2014) P10035}.

\bibitem{DWBC14} 
J. De Nardis, B. Wouters, M. Brockmann, and J.-S. Caux, 
\href{http://dx.doi.org/10.1103/PhysRevA.89.033601}{Phys. Rev. A {\bf 89}, 33601 (2014)}.

\bibitem{WDBF14} 
B. Wouters, J. De Nardis, M. Brockmann, D. Fioretto, M. Rigol, and J.-S. Caux, 
\href{http://dx.doi.org/10.1103/PhysRevLett.113.117202}{Phys. Rev. Lett. {\bf 113}, 117202 (2014)};\\
M. Brockmann, B. Wouters, D. Fioretto, J. De Nardis, R. Vlijm, and J.-S. Caux, 
\href{http://dx.doi.org/10.1088/1742-5468/2014/12/P12009}{J. Stat. Mech. (2014) P12009}.

\bibitem{PMWK14} 
B. Pozsgay, M. Mesty\'an, M. A. Werner, M. Kormos, G. Zar\'and, and G. Tak\'acs, 
\href{http://dx.doi.org/10.1103/PhysRevLett.113.117203}{Phys. Rev. Lett. {\bf 113}, 117203 (2014)};\\
M. Mesty\'an, B. Pozsgay, G. Tak\'acs, and M. A. Werner, 
\href{http://dx.doi.org/10.1088/1742-5468/2015/04/P04001}{J. Stat. Mech. (2015) P04001}.

\bibitem{BePC16} 
B. Bertini, L. Piroli, and P. Calabrese, 
\href{http://dx.doi.org/10.1088/1742-5468/2016/06/063102}{J. Stat. Mech. (2016) 63102};\\
M. Mesty\'{a}n, B. Bertini, L. Piroli, and P. Calabrese, 
\href{http://dx.doi.org/10.1088/1742-5468/aa7df0}{J. Stat. Mech. (2017) 83103}.

\bibitem{Bucc16} 
L. Bucciantini, 
\href{http://dx.doi.org/10.1007/s10955-016-1535-7}{J. Stat. Phys. {\bf 164}, 621 (2016)}.

\bibitem{PiCE16} 
L. Piroli, P. Calabrese, and F. H. L. Essler, 
\href{http://dx.doi.org/10.1103/PhysRevLett.116.070408}{Phys. Rev. Lett. {\bf 116}, 70408 (2016)};\\
L. Piroli, P. Calabrese, and F. H. L. Essler, 
\href{http://dx.doi.org/10.21468/SciPostPhys.1.1.001}{SciPost Phys. {\bf 1}, 1 (2016)}.

\bibitem{AlCa16_QA} 
V. Alba and P. Calabrese, 
\href{http://dx.doi.org/10.1088/1742-5468/2016/04/043105}{J. Stat. Mech. (2016) 043105}.

\bibitem{BeTC17} 
B. Bertini, E. Tartaglia, and P. Calabrese, 
\href{http://dx.doi.org/10.1088/1742-5468/aa8c2c}{J. Stat. Mech. (2017) 103107};\\
B. Bertini, E. Tartaglia, and P. Calabrese, 
\href{http://arxiv.org/abs/1802.10589}{arXiv:1802.10589 (2018)}.


\bibitem{cazalilla-06} M. A. Cazalilla, 
\href{http://dx.doi.org/10.1103/PhysRevLett.97.156403}{Phys. Rev. Lett. {\bf 97}, 156403 (2006)};\\
A. Iucci and M. A. Cazalilla. 
\href{http://dx.doi.org/10.1103/PhysRevA.80.063619}{Phys. Rev. A {\bf 80}, 063619 (2009)};\\
A. Iucci and M. A. Cazalilla. 
\href{http://dx.doi.org/10.1088/1367-2630/12/5/055019}{New J. Phys. {\bf 12}, 055019 (2010)}.

\bibitem{bpgd-09}
P. Barmettler, M. Punk, V. Gritsev, E. Demler, and E. Altman, 
\href{http://dx.doi.org/10.1103/PhysRevLett.102.130603}{Phys. Rev. Lett. {\bf 102}, 130603 (2009)};\\
P. Barmettler, M. Punk, V. Gritsev, E. Demler, and E. Altman, 
\href{http://dx.doi.org/10.1088/1367-2630/12/5/055017}{New J. Phys. {\bf 12}, 055017 (2010)};\\
V. Gritsev, T. Rostunov, and E. Demler, 
\href{http://dx.doi.org/10.1088/1742-5468/2010/05/P05012}{J. Stat. Mech. (2010) P05012}.

\bibitem{mc-12}
J. Mossel and J.-S. Caux, 
\href{http://dx.doi.org/10.1088/1367-2630/14/7/075006}{New J. Phys. {\bf 14}, 075006 (2012)}.

\bibitem{se-12}
D. Schuricht and F. H. L. Essler, 
\href{http://dx.doi.org/10.1088/1742-5468/2012/04/P04017}{J. Stat. Mech. (2012) P04017}.

\bibitem{csc-13} M. Collura, S. Sotiriadis, and P. Calabrese, 
\href{http://dx.doi.org/10.1103/PhysRevLett.110.245301}{Phys. Rev. Lett. {\bf 110}, 245301 (2013)};\\
M. Collura, S. Sotiriadis, and P. Calabrese, 
\href{http://dx.doi.org/10.1088/1742-5468/2013/09/P09025}{J. Stat. Mech. (2013) P09025}.

\bibitem{KoCC14} 
M. Kormos, M. Collura, and P. Calabrese, 
\href{http://dx.doi.org/10.1103/PhysRevA.89.013609}{Phys. Rev. A {\bf 89}, 13609 (2014)};\\
P. P. Mazza, M. Collura, M. Kormos, and P. Calabrese, 
\href{http://dx.doi.org/10.1088/1742-5468/2014/11/P11016}{J. Stat. Mech. (2014) P11016};\\
M. Collura, M. Kormos, and P. Calabrese, 
\href{http://arxiv.org/abs/1710.11615}{arXiv:1710.11615 (2017)}.

\bibitem{rs-14} 
M. A. Rajabpour and S. Sotiriadis, 
\href{http://dx.doi.org/10.1103/PhysRevA.89.033620}{Phys. Rev. A {\bf 89}, 033620 (2014)};\\
M. A. Rajabpour and S. Sotiriadis, 
\href{http://dx.doi.org/10.1103/PhysRevB.91.045131}{Phys. Rev. B {\bf 91}, 045131 (2015)}.

\bibitem{bkc-14} L. Bucciantini, M. Kormos, and P. Calabrese, 
\href{http://dx.doi.org/10.1007/s10955-016-1535-7}{J. Phys. A: Math. Theor. {\bf 47}, 175002 (2014)}.

\bibitem{dc-14}
J. De Nardis and J.-S. Caux, 
\href{http://dx.doi.org/10.1088/1742-5468/2014/12/P12012}{J. Stat. Mech. (2014) P12012}.

\bibitem{msca-16}
P. P. Mazza, J.-M. St\'{e}phan, E. Canovi, V. Alba, M. Brockmann, and M. Haque, 
\href{http://dx.doi.org/10.1088/1742-5468/2016/01/013104}{J. Stat. Mech. (2016) 013104}.

\bibitem{bf-16}
B. Bertini and M. Fagotti, 
\href{http://dx.doi.org/10.1103/PhysRevLett.117.130402}{Phys. Rev. Lett. {\bf 117}, 130402 (2016)}.


\bibitem{PiCa17} 
L. Piroli and P. Calabrese, 
\href{http://dx.doi.org/10.1103/PhysRevA.96.023611}{Phys. Rev. A {\bf 96}, 23611 (2017)}.


\bibitem{ia-12}
D. Iyer and N. Andrei, 
\href{http://dx.doi.org/10.1103/PhysRevLett.109.115304}{Phys. Rev. Lett. {\bf 109}, 115304 (2012)};\\
D. Iyer, H. Guan, and N. Andrei, 
\href{http://dx.doi.org/10.1103/PhysRevA.87.053628}{Phys. Rev. A {\bf 87}, 053628 (2013)};\\
G. Goldstein and N. Andrei, 
\href{http://arxiv.org/abs/1309.3471}{arXiv:1309.3471 (2013)};\\
W. Liu and N. Andrei, 
\href{http://dx.doi.org/10.1103/PhysRevLett.112.257204}{Phys. Rev. Lett. {\bf 112}, 257204 (2014)}.

\bibitem{mussardo-13}
G. Mussardo, 
\href{http://dx.doi.org/10.1103/PhysRevLett.111.100401}{Phys. Rev. Lett. {\bf 111}, 100401 (2013)}.

\bibitem{delfino-14}
G. Delfino, 
\href{http://dx.doi.org/10.1088/1751-8113/47/40/402001}{J. Phys. A: Math. Theor. {\bf 47}, 402001 (2014)};\\
G. Delfino and J. Viti, 
\href{http://arxiv.org/abs/1608.07612}{arXiv:1608.07612 (2016)}.

\bibitem{dpc-15}
J. De Nardis, L. Piroli, and J.-S. Caux, 
\href{http://dx.doi.org/10.1088/1751-8113/48/43/43FT01}{J. Phys. A: Math. Theor. {\bf 48}, 43FT01 (2015)}.

\bibitem{pe-16}
B. Pozsgay and V. Eisler, 
\href{http://dx.doi.org/10.1088/1742-5468/2016/05/053107}{J. Stat. Mech. (2016) 053107}.

\bibitem{cubero-16}
A. Cort\'{e}s Cubero, 
\href{http://dx.doi.org/10.1088/1742-5468/2016/08/083107}{J. Stat. Mech. (2016) 083107}.

\bibitem{vwed-16}
R. van den Berg, B. Wouters, S. Eli\"ens, J. De Nardis, R. M. Konik, and J.-S. Caux, 
\href{http://dx.doi.org/10.1103/PhysRevLett.116.225302}{Phys. Rev. Lett. {\bf 116}, 225302 (2016)}.

\bibitem{CuSc17} 
A. Cort\'es Cubero and D. Schuricht, 
\href{http://dx.doi.org/10.1088/1742-5468/aa8c2e}{J. Stat. Mech. (2017) 103106}.

\bibitem{Delf17} 
G. Delfino, 
\href{http://arxiv.org/abs/1710.06275}{arXiv:1710.06275 (2017)}.


\bibitem{kz-16}
M. Kormos and G. Zar\'{a}nd, 
\href{http://dx.doi.org/10.1103/PhysRevE.93.062101}{Phys. Rev. E {\bf 93}, 062101 (2016)};\\
C. P. Moca, M. Kormos, and G. Zar\'and, 
\href{http://arxiv.org/abs/1609.00974}{arXiv:1609.00974 (2016)}.


\bibitem{AlCa17} 
V. Alba and P. Calabrese, 
\href{http://dx.doi.org/10.1073/pnas.1703516114}{PNAS {\bf 114}, 7947 (2017)};\\
V. Alba and P. Calabrese, 
\href{http://arxiv.org/abs/1712.07529}{arXiv:1712.07529 (2017)}.





\bibitem{Pozs13} 
B. Pozsgay, 
\href{http://dx.doi.org/10.1088/1742-5468/2013/10/P10028}{J. Stat. Mech. (2013) P10028}.

\bibitem{PiPV17} 
L. Piroli, B. Pozsgay, and E. Vernier, 
\href{http://dx.doi.org/10.1088/1742-5468/aa5d1e}{J. Stat. Mech. (2017) 023106}.


\bibitem{LeUP98} 
P. R. Levstein, G. Usaj, and H. M. Pastawski, 
\href{http://dx.doi.org/10.1063/1.475664}{J. Chem. Phys. {\bf 108}, 2718 (1998)}.

\bibitem{PLUR00} 
H. M. Pastawski, P. R. Levstein, G. Usaj, J. Raya, and J. Hirschinger, 
\href{http://dx.doi.org/10.1016/S0378-4371(00)00146-1}{Physica A: Statist. Mech. Appl. {\bf 283}, 166 (2000)}.


\bibitem{qslz-06}
H. T. Quan, Z. Song, X. F. Liu, P. Zanardi, and C. P. Sun, 
\href{http://dx.doi.org/10.1103/PhysRevLett.96.140604}{Phys. Rev. Lett. {\bf 96}, 140604 (2006)};\\
L. Campos Venuti, N. T. Jacobson, S. Santra, and P. Zanardi, 
\href{http://dx.doi.org/10.1103/PhysRevLett.107.010403}{Phys. Rev. Lett. {\bf 107}, 010403 (2011)}.

\bibitem{silva-08}
A. Silva, 
\href{http://dx.doi.org/10.1103/PhysRevLett.101.120603}{Phys. Rev. Lett. {\bf 101}, 120603 (2008)};\\
A. Gambassi and A. Silva, 
\href{http://dx.doi.org/10.1103/PhysRevLett.109.250602}{Phys. Rev. Lett. {\bf 109}, 250602 (2012)};\\
S. Sotiriadis, A. Gambassi, and A. Silva, 
\href{http://dx.doi.org/10.1103/PhysRevE.87.052129}{Phys. Rev. E {\bf 87}, 052129 (2013)}.


\bibitem{pmgm-10}
F. Pollmann, S. Mukerjee, A. G. Green, and J. E. Moore, 
\href{http://dx.doi.org/10.1103/PhysRevE.81.020101}{Phys. Rev. E {\bf 81}, 020101 (2010)}.

\bibitem{fagotti2-13}
M. Fagotti, 
\href{http://arxiv.org/abs/1308.0277}{arXiv:1308.0277 (2013)}.

\bibitem{dpfz-13}
B. D\'{o}ra, F. Pollmann, J. Fort\'{a}gh, and G. Zar\'{a}nd, 
\href{http://dx.doi.org/10.1103/PhysRevLett.111.046402}{Phys. Rev. Lett. {\bf 111}, 046402 (2013)}.

\bibitem{hpk-13}
M. Heyl, A. Polkovnikov, and S. Kehrein, 
\href{http://dx.doi.org/10.1103/PhysRevLett.110.135704}{Phys. Rev. Lett. {\bf 110}, 135704 (2013)}.

\bibitem{ks-13}
C. Karrasch and D. Schuricht, 
\href{http://dx.doi.org/10.1103/PhysRevB.87.195104}{Phys. Rev. B {\bf 87}, 195104 (2013)}.

\bibitem{cwe-14}
E. Canovi, P. Werner, and M. Eckstein, 
\href{http://dx.doi.org/10.1103/PhysRevLett.113.265702}{Phys. Rev. Lett. {\bf 113}, 265702 (2014)}.

\bibitem{heyl-14}
M. Heyl, 
\href{http://dx.doi.org/10.1103/PhysRevLett.113.205701}{Phys. Rev. Lett. {\bf 113}, 205701 (2014)}.

\bibitem{vths-13}
R. Vasseur, K. Trinh, S. Haas, and H. Saleur, 
\href{http://dx.doi.org/10.1103/PhysRevLett.110.240601}{Phys. Rev. Lett. {\bf 110}, 240601 (2013)};\\
D. M. Kennes, V. Meden, and R. Vasseur, 
\href{http://dx.doi.org/10.1103/PhysRevB.90.115101}{Phys. Rev. B {\bf 90}, 115101 (2014)}.

\bibitem{as-14}
F. Andraschko and J. Sirker, 
\href{http://dx.doi.org/10.1103/PhysRevB.89.125120}{Phys. Rev. B {\bf 89}, 125120 (2014)}.

\bibitem{deluca-14}
A. De Luca, 
\href{http://dx.doi.org/10.1103/PhysRevB.90.081403}{Phys. Rev. B {\bf 90}, 081403 (2014)}.

\bibitem{heyl-15}
M. Heyl, 
\href{http://dx.doi.org/10.1103/PhysRevLett.115.140602}{Phys. Rev. Lett. {\bf 115}, 140602 (2015)};\\
M. Heyl, 
\href{http://arxiv.org/abs/1608.06659}{arXiv:1608.06659 (2016)}.

\bibitem{kk-14}
J. N. Kriel, C. Karrasch, and S. Kehrein, 
\href{http://dx.doi.org/10.1103/PhysRevB.90.125106}{Phys. Rev. B {\bf 90}, 125106 (2014)}.

\bibitem{vd-14}
S. Vajna and B. D\'{o}ra, 
\href{http://dx.doi.org/10.1103/PhysRevB.89.161105}{Phys. Rev. B {\bf 89}, 161105 (2014)}.

\bibitem{ps-14}
T. P\'{a}lmai and S. Sotiriadis, 
\href{http://dx.doi.org/10.1103/PhysRevE.90.052102}{Phys. Rev. E {\bf 90}, 052102 (2014)};\\
T. Palmai, 
\href{http://dx.doi.org/10.1103/PhysRevB.92.235433}{Phys. Rev. B {\bf 92}, 235433 (2015)}.

\bibitem{ssd-15}
S. Sharma, S. Suzuki, and A. Dutta, 
\href{http://dx.doi.org/10.1103/PhysRevB.92.104306}{Phys. Rev. B {\bf 92}, 104306 (2015)}.

\bibitem{sdpd-16}
U. Divakaran, S. Sharma, and A. Dutta, 
\href{http://dx.doi.org/10.1103/PhysRevE.93.052133}{Phys. Rev. E {\bf 93}, 052133 (2016)};\\
S. Sharma, U. Divakaran, A. Polkovnikov, and A. Dutta, 
\href{http://dx.doi.org/10.1103/PhysRevB.93.144306}{Phys. Rev. B {\bf 93}, 144306 (2016)}.

\bibitem{zhks-16}
B. Zunkovic, A. Silva, and M. Fabrizio, 
\href{http://dx.doi.org/10.1098/rsta.2015.0160}{Phil. Trans. R. Soc. A {\bf 374}, 20150160 (2016)};\\
B. Zunkovic, M. Heyl, M. Knap, and A. Silva, 
\href{http://arxiv.org/abs/1609.08482}{arXiv:1609.08482 (2016)}.

\bibitem{zy-16}
J. M. Zhang and H.-T. Yang, 
\href{http://dx.doi.org/10.1209/0295-5075/114/60001}{EPL {\bf 114}, 60001 (2016)};\\
J. M. Zhang and H.-T. Yang, 
\href{http://dx.doi.org/10.1209/0295-5075/116/10008}{EPL {\bf 116}, 10008 (2016)}.

\bibitem{ps-16}
T. Puskarov and D. Schuricht, 
\href{http://dx.doi.org/10.21468/SciPostPhys.1.1.003}{SciPost Phys. {\bf 1}, 003 (2016)}.

\bibitem{Heyl17} 
M. Heyl, 
\href{http://arxiv.org/abs/1709.07461}{arXiv:1709.07461 (2017)}.


\bibitem{JaJo17} 
R. Jafari and H. Johannesson, 
\href{http://dx.doi.org/10.1103/PhysRevLett.118.015701}{Phys. Rev. Lett. {\bf 118}, 015701 (2017)}.



\bibitem{sd-11}
J.-M. St\'{e}phan and J. Dubail, 
\href{http://dx.doi.org/10.1088/1742-5468/2011/08/P08019}{J. Stat. Mech. (2011) P08019}.

\bibitem{Card14}
J. Cardy, 
\href{http://dx.doi.org/10.1103/PhysRevLett.112.220401}{Phys. Rev. Lett. {\bf 112}, 220401 (2014)};\\
J. Cardy, 
\href{http://dx.doi.org/10.1088/1742-5468/2016/02/023103}{J. Stat. Mech. (2016) 23103};\\
J. Cardy, 
\href{http://dx.doi.org/10.1088/1751-8113/49/41/415401}{J. Phys. A: Math. Theor. {\bf 49}, 415401 (2016)}.

\bibitem{Step17} 
J.-M. St\'ephan, 
\href{http://dx.doi.org/10.1088/1742-5468/aa8c19}{J. Stat. Mech. (2017) 103108}.

\bibitem{NaRa17} 
K. Najafi and M. A. Rajabpour, 
\href{http://dx.doi.org/10.1103/PhysRevB.96.014305}{Phys. Rev. B {\bf 96}, 014305 (2017)}.

\bibitem{KrLM17} 
P. L. Krapivsky, J. M. Luck, and K. Mallick, 
\href{http://arxiv.org/abs/1710.08178}{arXiv:1710.08178 (2017)}.


\bibitem{klum-93} A. Kl\"{u}mper, \href{http://dx.doi.org/doi:10.1007/BF01316831}{Z. Physik B {\bf 91}, 507 (1993)}.

\bibitem{suzuki-99}
J. Suzuki, 
\href{http://dx.doi.org/10.1088/0305-4470/32/12/008}{J. Phys. A: Math. Gen. {\bf 32}, 2341 (1999)}.

\bibitem{klumper-04} 
A. Kl\"{u}mper, 
\href{http://dx.doi.org/10.1007/BFb0119598}{Lect. Notes Phys. {\bf 645}, 349 (2004)}


\bibitem{PiPV17_int} 
L. Piroli, B. Pozsgay, and E. Vernier, 
\href{http://dx.doi.org/10.1016/j.nuclphysb.2017.10.012}{Nucl. Phys. B {\bf 925}, 362 (2017)}.

\bibitem{BrSt17} 
M. Brockmann and J.-M. St\'{e}phan, 
\href{http://dx.doi.org/10.1088/1751-8121/aa809c}{J. Phys. A: Math. Theor. {\bf 50}, 354001 (2017)}.

\bibitem{kbi-93} V.E. Korepin, N.M. Bogoliubov and A.G. Izergin, 
{\it Quantum inverse scattering method and correlation functions}, Cambridge University Press (1993);\\
F. H. L. Essler, H. Frahm, F. G\"ohmann, A. Kl\"umper, and V. E. Korepin,  
{\it The One-Dimensional Hubbard Model}, Cambridge University Press (2005).


\bibitem{skly-88}
E. K. Sklyanin, 
\href{http://dx.doi.org/10.1088/0305-4470/21/10/015}{J. Phys. A: Math. Gen. {\bf 21}, 2375 (1988)}.

\bibitem{kkmn-07}
N. Kitanine, K. K. Kozlowski, J. M. Maillet, G. Niccoli, N. A. Slavnov, and V. Terras, 
\href{http://dx.doi.org/10.1088/1742-5468/2007/10/P10009}{J. Stat. Mech. (2007) P10009};\\
N. Kitanine, K. K. Kozlowski, J. M. Maillet, G. Niccoli, N. A. Slavnov, and V. Terras, 
\href{http://dx.doi.org/10.1088/1742-5468/2008/07/P07010}{J. Stat. Mech. (2008) P07010}.

\bibitem{kmn-14}
N. Kitanine, J. M. Maillet, and G. Niccoli, 
\href{http://dx.doi.org/10.1088/1742-5468/2014/05/P05015}{J. Stat. Mech. (2014) P05015}.

\bibitem{wycs-15}
Y. Wang,  W.-L. Yang, J. Cao, K. Shi,
{\it Off-diagonal Bethe ansatz for exactly solvable models}, Springer (2015). 




\bibitem{nepomechie-02}
R. I. Nepomechie, 
\href{http://dx.doi.org/10.1016/S0550-3213(01)00585-5}{Nucl. Phys. B {\bf 622}, 615 (2002)}.

\bibitem{nepomechie-04}
R. I. Nepomechie, 
\href{http://dx.doi.org/10.1088/0305-4470/37/2/012}{J. Phys. A: Math. Gen. {\bf 37}, 433 (2004)}.


\bibitem{clsw-03}
J. Cao, H.-Q. Lin, K.-J. Shi, and Y. Wang, 
\href{http://dx.doi.org/10.1016/S0550-3213(03)00372-9}{Nucl. Phys. B {\bf 663}, 487 (2003)}.

\bibitem{fgsw-11}
H. Frahm, J. H. Grelik, A. Seel, and T. Wirth, 
\href{http://dx.doi.org/10.1088/1751-8113/44/1/015001}{J. Phys. A: Math. Theor. {\bf 44}, 015001 (2011)}.

\bibitem{niccoli-12}
G. Niccoli, 
\href{http://dx.doi.org/10.1088/1742-5468/2012/10/P10025}{J. Stat. Mech. (2012) P10025},\\
S. Faldella, N. Kitanine, and G. Niccoli, 
\href{http://dx.doi.org/10.1088/1742-5468/2014/01/P01011}{J. Stat. Mech. (2014) P01011}.

\bibitem{cysw-13}
J. Cao, W.-L. Yang, K. Shi, and Y. Wang, 
\href{http://dx.doi.org/10.1103/PhysRevLett.111.137201}{Phys. Rev. Lett. {\bf 111}, 137201 (2013)};\\
J. Cao, W.-L. Yang, K. Shi, and Y. Wang, 
\href{http://dx.doi.org/10.1016/j.nuclphysb.2013.10.001}{Nucl. Phys. B {\bf 877}, 152 (2013)};\\
J. Cao, W.-L. Yang, K. Shi, and Y. Wang, 
\href{http://dx.doi.org/10.1016/j.nuclphysb.2013.06.022}{Nucl. Phys. B {\bf 875}, 152 (2013)};\\
J. Cao, W.-L. Yang, K. Shi, and Y. Wang, 
\href{http://dx.doi.org/10.1088/1751-8113/48/44/444001}{J. Phys. A: Math. Theor. {\bf 48}, 444001 (2015)}.

\bibitem{nepomechie-13}
R. I. Nepomechie, 
\href{http://dx.doi.org/10.1088/1751-8113/46/44/442002}{J. Phys. A: Math. Theor. {\bf 46}, 442002 (2013)}.

\bibitem{NepoWang14}  
R. I. Nepomechie, C. Wang, 
\href{http://dx.doi.org/10.1088/1751-8113/47/3/032001}{J. Phys. A: Math. Theor. {\bf 47} 032001 (2014)}.


\bibitem{kns-11} 
A. Kuniba, T. Nakanishi, and J. Suzuki, 
\href{http://dx.doi.org/10.1088/1751-8113/44/10/103001}{J. Phys. A: Math. Theor. {\bf 44}, 103001 (2011)}.


\bibitem{zhou-95}
Y. Zhou, 
\href{http://dx.doi.org/10.1016/0550-3213(95)00293-2}{Nucl. Phys. B {\bf 453}, 619 (1995)}.

\bibitem{mn-92}
L. Mezincescu and R. I. Nepomechie, 
\href{http://dx.doi.org/10.1088/0305-4470/25/9/024}{J. Phys. A: Math. Gen. {\bf 25}, 2533 (1992)}.

\bibitem{zhou-96}
Y. Zhou, 
\href{http://dx.doi.org/10.1016/0550-3213(95)00553-6}{Nucl. Phys. B {\bf 458}, 504 (1996)}.

\bibitem{Kuniba} 
A. Kuniba, K. Sakai, and J. Suzuki, 
\href{http://dx.doi.org/10.1016/S0550-3213(98)00300-9}{Nucl. Phys. B {\bf 525}, 597 (1998)}.

\bibitem{BeSt99} 
A. Belavin and Y. Stroganov, 
\href{http://dx.doi.org/10.1016/S0370-2693(99)01150-8}{Phys. Lett. B {\bf 466}, 281 (1999)}.

\bibitem{BeGF00} 
A. A. Belavin, S. Y. Gubanov, and B. L. Feigin, 
\href{http://arxiv.org/abs/hep-th/0008011}{arxiv:Hep-Th/0008011 (2000)}.

\bibitem{Nepo03} 
R. I. Nepomechie, 
\href{http://dx.doi.org/10.1023/A:1023016602955}{J. Stat. Phys. {\bf 111}, 1363 (2003)};\\
R. Murgan, R. I. Nepomechie, and C. Shi, 
\href{http://dx.doi.org/10.1088/1742-5468/2006/08/P08006}{J. Stat. Mech. (2006) P08006}.


\bibitem{FGSW11} 
H. Frahm, J. H. Grelik, A. Seel, and T. Wirth, 
\href{http://dx.doi.org/10.1088/1751-8113/44/1/015001}{J. Phys. A: Math. Theor. {\bf 44}, 015001 (2011)}.



\bibitem{KlPe93} 
A. Kl\"{u}mper and P. A. Pearce, 
\href{http://dx.doi.org/10.1016/0378-4371(93)90371-A}{Physica A: Stat. Mech. Appl. {\bf 194}, 397 (1993)}.


\bibitem{JuKS98} 
G. J\"{u}ttner, A. Kl\"{u}mper, and J. Suzuki, 
\href{http://dx.doi.org/10.1016/S0550-3213(97)00772-4}{Nucl. Phys. B {\bf 512}, 581 (1998)}.


\bibitem{DoTa96} 
P. Dorey and R. Tateo, 
\href{http://dx.doi.org/10.1016/S0550-3213(96)00516-0}{Nucl. Phys. B {\bf 482}, 639 (1996)}.

\bibitem{BaLZ97} 
V. V. Bazhanov, S. L. Lukyanov, and A. B. Zamolodchikov, 
\href{http://dx.doi.org/10.1016/S0550-3213(97)00022-9}{Nucl. Phys. B {\bf 489}, 487 (1997)}.



\bibitem{Vida07} 
G. Vidal, 
\href{http://dx.doi.org/10.1103/PhysRevLett.98.070201}{Phys. Rev. Lett. {\bf 98}, 070201 (2007)}.








\end{thebibliography}
\end{document}